\documentclass[12pt]{article}

\def\square{\kern1pt\vbox{\hrule height 1.2pt\hbox{\vrule width 1.2pt\hskip 3pt
   \vbox{\vskip 6pt}\hskip 3pt\vrule width 0.6pt}\hrule height 0.6pt}\kern1pt}

\begin{document}

\begin{titlepage}

\begin{flushright}
UFIFT-QG-13-01
\end{flushright}

\begin{center}
{\bf One Loop Field Strengths of Charges and Dipoles on a Locally de
Sitter Background}
\end{center}

\begin{center}
H. Degueldre$^{\dagger}$
\end{center}

\begin{center}
\it{Department of Nonlinear Dynamics \\
Max-Planck-Institute for Dynamics and Self-Organization \\
37077 Goettingen, GERMANY}
\end{center}

\begin{center}
and
\end{center}

\begin{center}
R. P. Woodard$^{\ddagger}$
\end{center}

\begin{center}
\it{Department of Physics \\
University of Florida \\
Gainesville, FL 32611 USA}
\end{center}

\begin{center}
ABSTRACT
\end{center}
We use the one loop vacuum polarization induced by scalar quantum
electrodynamics to compute the electric and magnetic fields of point
charges and magnetic dipoles on a locally de Sitter background. Our
results are consistent with the physical picture of an inflating
universe filling with a vast sea of charged particles as more and
more virtual infrared scalar are ripped out of the vacuum. One
consequence is that vacuum polarization quickly becomes nonperturbatively
strong. Our computation employs the Schwinger-Keldysh effective field
equations and is done in flat, conformal coordinates. Results are also
obtained for static coordinates.

\begin{flushleft}
\end{flushleft}

\begin{flushleft}
$^{\dagger}$ e-mail: hdegueld@nld.ds.mpg.de \\
$^{\ddagger}$ e-mail: woodard@phys.ufl.edu
\end{flushleft}
\end{titlepage}

\section{Introduction}

The phenomenon of vacuum polarization is a triumph of flat space
Quantum Electrodynamics (QED). The enhanced high energy coupling
strength it predicts has not only been verified experimentally, it also
provided the crucial paradigm for understanding renormalization group
flows. In spite of these successes, the enhancement is only about 2\%
at the highest energy accelerators we have so far been able to build
\cite{Peskin}.

Loop effects such as vacuum polarization derive ultimately from the
response to virtual particles (in this case, electrons and positrons)
which are the quantum field theoretic manifestation of 0-point motion.
It has long been realized that these effects must be strengthened in
an expanding universe, essentially because the expansion of spacetime
tends to hold virtual quanta apart \cite{Schr}. Leonard Parker made
the first quantitative computations in the late 60's \cite{Parker1}.
He found that the effect is largest for massless particles that are not
conformally invariant \cite{Parker2}, which includes massless, minimally
coupled scalars and gravitons \cite{Grishchuk}. At a fixed expansion rate
the effect increases with the cosmological acceleration. During primordial
inflation massless virtual quanta are actually ripped out of the vacuum,
which is thought to be the origin of primordial scalar and tensor
perturbations \cite{SMC}.

From this discussion one can infer that the largest possible vacuum
polarization occurs during de Sitter inflation (which has the highest
acceleration consistent with stability) and derives from massless
charged particles that are not conformally invariant. These conditions
are realistic. The measured value of the scalar power spectrum, and the
current upper bound  on the tensor-to-scalar ratio \cite{WMAP,SPT} suggest
that the deceleration parameter (which is minus the acceleration) of
primordial inflation was less than -.993 \cite{KOW1}, which amply justifies
taking the de Sitter limit of -1. Primordial inflation is thought to have
occurred at such an enormous scale that all known charged particles would
have been effectively massless. However, massless fermions are conformally
invariant on the classical level, which means they can only experience the
expansion of spacetime through the conformal anomaly. This is responsible
for the vacuum polarization of ordinary QED being only slightly enhanced
during primordial inflation \cite{Dolgov}.

Much larger effects can come from a charged scalar (such as the components
of the Higgs which become the longitudinal parts of the $W^{\pm}$ in low
energy physics) provided it is massless and not conformally invariant.
Davis, Dimopoulos, Prokopec and T\"ornkvist made the remarkable proposal
that such a particle might even endow the photon with mass during
primordial inflation \cite{DDPT,PW1}. Their idea was confirmed with a
dimensionally regulated and fully renormalized one loop computation of
the vacuum polarization from Scalar Quantum Electrodynamics (SQED) on de
Sitter background \cite{PTW}. Although the one loop effect grows without
bound \cite{PW2}, a nonperturbative resummation of the leading secular
terms reveals that the photon mass approaches a constant value of about
1.8163 times the Hubble constant \cite{PTsW}.

The quantum-corrected, linearized Maxwell equations in an arbitrary
metric $g_{\mu\nu}$ read,
\begin{equation}
\partial_{\nu} \Bigl[ \sqrt{-g} \, g^{\nu\rho} g^{\mu\sigma}
F_{\rho\sigma}(x)\Bigr] + \int \!\! d^4x' \Bigl[\mbox{}^{\mu}
\Pi^{\nu}\Bigr](x;x') A_{\nu}(x') = J^{\mu}(x) \; . \label{effeqn}
\end{equation}
Here $A_{\mu}(x)$ stands for the vector potential,
$[\mbox{}^{\mu}\Pi^{\nu}](x;x')$ for the vacuum polarization and
$J^{\mu}(x)$ is the current density. It is immediately obvious that
the same vacuum polarization that reveals the effective mass of
dynamical photons can also be used to compute the electrodynamic
response to point charges and current dipoles. That is the purpose
of this paper.

Throughout this paper we employ a spacelike metric. Section 2 derives
corrections to the field strengths of point charges and dipoles from
the one loop vacuum polarization of flat space SQED. That serves as a
useful correspondence limit and also illustrates the basics of the
more challenging de Sitter computation. In section 3 we review the de
Sitter geometry. Section 4 presents the actual de Sitter computation,
with the messy details consigned to two appendices. Our conclusions
comprise section 5.

\section{Vacuum Polarization in Flat Space}

The purpose of this section is to work out the effect at one loop
from a massless, charged scalar in flat space. Our analysis parallels
recent computations, made using the Schwinger-Keldysh formalism
\cite{earlySK,lateSK}, of the one loop effect of gravitons on
electromagnetism \cite{LW} and of a massless, minimally coupled
scalar on linearized gravity in flat space background \cite{SPW}.
We begin with some considerations following from the general tensor
structure of the vacuum polarization and from the fact that we only
have the result to some finite order in the loop expansion. Then the
one loop correction is derived for a point charge and for a point
magnetic dipole.

\subsection{General considerations}

Poincar\'e invariance and transversality constrain the vacuum polarization
at any order to take the form,
\begin{equation}
\Bigl[ \mbox{}^{\mu} \Pi^{\nu}\Bigr](x;x') = \Bigl[ \partial' \!\cdot\!
\partial \, \eta^{\mu\nu} \!-\! \partial^{\prime \mu} \partial^{\nu} \Bigr]
\Pi(x \!-\! x') = -\Bigl[ \partial' \!\cdot\! \partial' \, \eta^{\mu\nu}
\!-\! \partial^{\prime \mu} \partial^{\prime \nu} \Bigr] \Pi(x \!-\! x')
\; . \label{flatform}
\end{equation}
This means we can integrate by parts to express the effective field
equation (\ref{effeqn}) as,
\begin{equation}
\partial_{\nu} F^{\nu\mu}(x) - \int \!\! d^4x' \, \Pi(x \!-\! x')
\partial'_{\nu} F^{\nu\mu}(x') + \Bigl({\rm Surface\ Terms}\Bigr)
= J^{\mu}(x) \; . \label{flateqn0}
\end{equation}
The causality of the Schwinger-Keldysh formalism precludes there being
any spatial surface terms, or any surface terms at the upper limit of
the temporal integration \cite{TW1}. There can be surface terms at the
lower temporal limit, which corresponds to the time when the state was
released \cite{FW}. We shall assume that these are completely absorbed
by perturbative corrections to the initial state wave functional
\cite{KOW2}. If we agree to exclude these state corrections from
$\Pi(x \!-\! x')$, the linearized effective field equations become just,
\begin{equation}
\partial_{\nu} F^{\nu\mu}(x) - \int \!\! d^4x' \, \Pi(x \!-\! x')
\partial'_{\nu} F^{\nu\mu}(x') = J^{\mu}(x) \; . \label{flateqn1}
\end{equation}

The one loop result for $\Pi(x \!-\! x')$ is \cite{PTW},
\begin{equation}
\Pi^{(1)}(x \!-\! x') = -\frac{\alpha \partial^4}{96 \pi^2}
\Biggl[ \theta(\Delta t \!-\! \Delta x) \Biggl\{ \ln\Bigl[ \mu^2
(\Delta t^2 \!-\! \Delta x^2)\Bigr] \!-\! 1\Biggr\} \Biggr] \; .
\label{flatPi1}
\end{equation}
Here $\partial^2 \equiv \eta^{\mu\nu} \partial_{\mu} \partial_{\nu}
= -\partial_t^2 + \nabla^2$, $\Delta t \equiv t - t'$, $\Delta x
\equiv \Vert \vec{x} - \vec{x}' \Vert$, and $\alpha \equiv e^2/4\pi$
is the loop-counting parameter of SQED. Because we do not possess
the higher order corrections, equation (\ref{flateqn1}) can only be
solved perturbatively. That is, one expands the field strength in
powers of $\alpha$,
\begin{equation}
F_{\mu\nu} = F^{(0)}_{\mu\nu} + F^{(1)}_{\mu\nu} + F^{(2)}_{\mu\nu}
+ \dots \; ,
\end{equation}
and then distills (\ref{flateqn1}) into terms of the same order.
Because the current density is zeroth order we have,
\begin{eqnarray}
\partial^{\nu} F^{(0)}_{\nu\mu}(x) & = & J_{\mu}(x) \; , \\
\partial^{\nu} F^{(1)}_{\nu\mu}(x) & = & \int\!\! d^4x' \,
\Pi^{(1)}(x \!-\! x') \partial^{\prime \nu} F^{(0)}_{\nu\mu}(x')  \; ,
\label{flateqn2}
\end{eqnarray}
and so on. We can therefore define the source of the one loop
field strength to be the ``one loop current density,''
\begin{equation}
J^{(1)}_{\mu}(x) \equiv \int \!\! d^4x' \, \Pi^{(1)}(x \!-\! x')
J_{\mu}(x') \; . \label{J1}
\end{equation}

It remains to explain that we will always solve for the field
strength tensor directly, rather than going through the intermediate
step of finding the vector potential. Note that contracting the
Levi-Civita density into the gradient of $\partial^{\alpha}
F_{\alpha\nu}$ gives,
\begin{equation}
\epsilon^{\rho\sigma\mu\nu} \partial_{\mu} \partial^{\alpha}
F_{\alpha \nu} = \frac12 \partial^2 \epsilon^{\rho\sigma\mu\nu}
F_{\mu\nu} \; . \label{neatrel}
\end{equation}
Substituting (\ref{J1}) into (\ref{flateqn2}) and then using
(\ref{neatrel}) implies,
\begin{equation}
\partial^2 F^{(1)}_{\mu\nu} = \partial_{\mu} J^{(1)}_{\nu}
- \partial_{\nu} J^{(1)}_{\mu} \label{flateqn3} \; .
\end{equation}
This form is advantageous in view of expression (\ref{flatPi1}) for
$\Pi^{(1)}(x - x')$. Up to a homogeneous solution, we can write the
one loop field strength tensor as $F^{(1)}_{\mu\nu} = \partial_{\mu}
\mathcal{J}^{(1)}_{\nu} - \partial_{\nu} \mathcal{J}^{(1)}_{\mu}$,
where $\mathcal{J}^{(1)}_{\mu}$ is obtained by simply removing a
factor of $\partial^2$ from $J^{(1)}_{\mu}$,
\begin{equation}
\mathcal{J}^{(1)}_{\mu}(x) = -\frac{\alpha \partial^2}{96 \pi^2}
\int \!\! d^4x' \, \theta(\Delta t \!-\! \Delta x) \Biggl\{
\ln\Bigl[ \mu^2 (\Delta t^2 \!-\! \Delta x^2)\Bigr] \!-\! 1 \Biggr\}
J_{\mu}(x') \; . \label{scriptJ}
\end{equation}
The ambiguity regarding homogeneous solutions can be settled by
appealing to initial conditions.

\subsection{Response to a point charge}

The current density of a stationary charge $q$ at the origin is,
\begin{equation}
J^{\mu}(t,\vec{x}) = q \delta^{\mu}_0 \delta^3( \vec{x}) \qquad
\Longleftrightarrow \qquad J_{\mu}(t,\vec{x}) = -q \delta^0_{\mu}
\delta^3(\vec{x}) \; . \label{charge}
\end{equation}
Taking the initial time to be $t_I$ and inserting (\ref{charge})
in expression (\ref{scriptJ}) gives,
\begin{eqnarray}
\lefteqn{\mathcal{J}^{(1)}_{\mu}(t,\vec{x}) = \frac{\alpha q
\delta^0_{\mu}}{96 \pi^2} \, \partial^2 \! \int_{t_I}^{t-x} \!\!\!\!\!
dt' \, \Biggl\{ \ln\Bigl[ \mu (t \!-\! t' \!-\! x)\Bigr] \!+\! \ln\Bigl[\mu
(t \!-\! t' \!+\! x)\Bigr] \!-\! 1\Biggr\} \; , } \\
& & = \frac{\alpha q \delta^0_{\mu}}{96 \pi^2} \, \partial^2 \Biggl\{
-2 x \ln(2 \mu x) \!+\! 3 (x \!-\! t \!+\! t_I) \!+\!
(t \!-\! t_I \!-\! x) \ln\Bigl[ \mu (t \!-\! t_I \!-\! x)\Bigr] \nonumber \\
& & \hspace{6.5cm} + (t \!-\! t_I \!+\! x) \ln\Bigl[ \mu (t \!-\! t_I
\!+\! x)\Bigr] \Biggr\} \; . \qquad \label{scriptJ2}
\end{eqnarray}
Before acting the d'Alembertian it is useful to specialize it to
functions of $t$ and $x$,
\begin{equation}
\partial^2 \longrightarrow -\partial_t^2 + \frac1{x} \, \partial_x^2 x
= \frac1{2 x} (\partial_t \!-\! \partial_x) (\partial_t \!+\! \partial_x)
(t \!-\! x) - \frac1{2 x} (\partial_t \!-\! \partial_x) (\partial_t \!+\!
\partial_x) (t \!+\! x) \; . \label{DAL}
\end{equation}
We now use (\ref{DAL}) to act the d'Alembertian in (\ref{scriptJ2}),
\begin{equation}
\mathcal{J}^{(1)}_{\mu}(t,\vec{x}) = \frac{\alpha q \delta^0_{\mu}}{96 \pi^2}
\Biggl\{ -\frac4{x} \, \ln(2 \mu x) + \frac2{x} \ln\Bigl[ \frac{t \!-\!
t_I \!+\! x}{t \!-\! t_I \!-\! x}\Bigr] \Biggr\} \; .
\end{equation}
The homogeneous terms would vanish if acted upon by another d'Almbertian,
but they also drop out when $t_I$ is taken to the infinite past,
\begin{equation}
\lim_{t_I \rightarrow -\infty} \mathcal{J}^{(1)}_{\mu}(t,\vec{x}) =
-\frac{\alpha q \delta^0_{\mu}}{24 \pi^2} \, \frac{\ln(2 \mu x)}{x} \; .
\label{scriptJ3}
\end{equation}
The one loop field strengths follow,
\begin{eqnarray}
F^{(1)}_{0i}(t,\vec{x}) & = & -\frac{\alpha q}{24 \pi^2} \, \frac{x^i}{x^3}
\Bigl[ \ln(2 \mu x) \!-\! 1\Bigr] = -\frac{\alpha}{6 \pi} \Bigl[ \ln(2 \mu x)
\!-\! 1\Bigr] \times F^{(0)}_{0i}(t,\vec{x}) \; , \qquad \label{Efield} \\
F^{(1)}_{ij}(t,\vec{x}) & = & 0 \label{nofield} \; .
\end{eqnarray}

From expression (\ref{Efield}) we see that the regularization scale
$\mu$ gives rise to a characteristic length, $L \equiv e^1/2\mu$.
Outside this length the charge $q$ is screened by quantum corrections,
whereas it is enhanced for $x < L$. Why this happens becomes clearer
from examining the one loop current density,
\begin{equation}
J^{\mu}_{(1)}(t,\vec{x}) = \partial^2 \mathcal{J}^{\mu}_{(1)}(t,\vec{x})
= \frac{\alpha q \delta^{\mu}_0}{24 \pi^2} \Biggl\{- \frac1{x^3} +
\infty \times \delta^3(\vec{x}) \Biggr\} \; . \label{1loopJ}
\end{equation}
Of course it is the positive contribution at the origin which enhances
the electrostatic force for $x < L$, and the negative cloud of charge
density screens it for $x > L$.

Vacuum polarization can be understood physically from the response of
virtual charged particles (in this case massless scalars) to the
classical source. The energy-time uncertainty principle allows a pair
of virtual particles with mass $m$ and wave number $k$ to exist for a
time $\Delta t \sim 1/\sqrt{m^2 + k^2}$. During this time the partner
whose charge is opposite to $q$ will be pulled into the source, and the
partner with the same charge as $q$ will be pushed away. By itself, that
would lead to a negative induced charge at the origin, which is exactly
opposite to (\ref{1loopJ}). However, two features complicate the
physics of vacuum polarization in this model: renormalization and the
masslessness of our scalars. The first feature means that what we call
the one loop current density $J_{\mu}^{(1)}(t,\vec{x})$ actually includes
an infinite constant times the classical current density. Had the mass
been nonzero, this infinite constant would have been chosen to completely
null the one loop correction to the fields at spatial infinity. However,
the masslessness of our charged scalars means that there continues to
be an effect out to arbitrarily large distances. In that case, one chooses
the renormalization constants to make the one loop field strengths vanish
at some fixed distance, $L = e^1/2\mu$.

\subsection{Response to a point magnetic dipole}

Shrinking a current loop with magnetic dipole moment $\vec{m}$ down to
the origin gives the following current density,
\begin{equation}
J^{\mu}(t,\vec{x}) = -\epsilon^{0 \mu\rho\sigma}
m_{\rho} \partial_{\sigma} \delta^3( \vec{x}) \quad \Longleftrightarrow
\quad J^0 = 0 \quad , \quad \vec{J} = -\vec{m} \!\times\! \vec{\nabla}
\delta^3(\vec{x}) \; . \label{dipole}
\end{equation}
This differs from the current density (\ref{charge}) of a point charge
only by the replacement,
\begin{equation}
q \delta^{\mu}_0 \longrightarrow -\epsilon^{0\mu\rho\sigma} m_{\rho}
\partial_{\sigma} \; . \label{replacement}
\end{equation}
Because the derivative must be spatial it can be partially integrated,
then converted from $\vec{\nabla}'$ to $-\vec{\nabla}$, so the result
for $\mathcal{J}_{(1)}^{\mu}(t,\vec{x})$ follows by making the same
replacement (\ref{replacement}) in expression (\ref{scriptJ3}),
\begin{equation}
\lim_{t_I \rightarrow -\infty} \mathcal{J}_{(1)}^{\mu}(t,\vec{x}) =
-\frac{\alpha}{24 \pi^2} \epsilon^{0\mu\rho\sigma} m_{\rho}
\partial_{\sigma} \Bigl[ \frac{\ln(2\mu x)}{x} \Bigr] \; . \label{COV}
\end{equation}
It is useful to 3 + 1 decompose (\ref{COV}),
\begin{eqnarray}
\mathcal{J}^{(1)}_0(t,\vec{x}) & = & 0 \; , \\
\mathcal{J}^{(1)}_j(t,\vec{x}) & = & \frac{\alpha}{6 \pi} \Bigl[
\ln(2 \mu x) \!-\! 1\Bigr] \times \frac{\epsilon^{jk\ell} m^k x^{\ell}}{
4 \pi x^3} \; . \qquad
\end{eqnarray}
The resulting field strengths are,
\begin{eqnarray}
F_{0i}(t,\vec{x}) & = & 0 \; , \\
F_{ij}(t,\vec{x}) & = & \frac{\alpha \epsilon^{ijk}}{24 \pi^2} \Biggl\{
\frac{(3 \widehat{x}^k \widehat{x} \!\cdot\! \vec{m} \!-\! m^k)}{x^3}
\Bigl[ \ln(2\mu x) \!-\! 1\Bigr] + \frac{(m^k \!-\! \widehat{x}^k
\widehat{x} \!\cdot\! \vec{m})}{x^3} \Biggr\} . \qquad \label{flatB}
\end{eqnarray}

In view of the relation $F_{ij} = -\epsilon^{ijk} B^k$ we can read off
the one loop correction to the magnetic field from (\ref{flatB}),
\begin{equation}
\vec{B}^{(1)}(t,\vec{x}) = -\frac{\alpha}{6\pi} \Bigl[\ln(2\mu x) \!-\!
1\Bigr] \times \vec{B}^{(0)}(t,\vec{x}) -\frac{\alpha}{6 \pi} \times
\frac{(\vec{m} \!-\! \widehat{x} \widehat{x} \!\cdot\! \vec{m})}{4\pi x^3}
\; , \label{1loopB}
\end{equation}
where (away from $\vec{x} = 0$) the classical result is,
\begin{equation}
\vec{B}^{(0)}(t,\vec{x}) = \frac{(3 \widehat{x} \widehat{x} \!\cdot\!
\vec{m} \!-\! \vec{m})}{4 \pi x^3} \; .
\end{equation}
Expression (\ref{1loopB}) represents the same sort of screening we found
in (\ref{Efield}) for a point charge. The second term in (\ref{1loopB})
is just the residue necessary to keep $\vec{B}$ transverse with the
coordinate-dependent screening factor. Except for the usual divergence
at the origin, the one loop current density which supports this field is,
\begin{equation}
\vec{J}^{(1)}(t,\vec{x}) = -\frac{\alpha}{8 \pi^2} \frac{\vec{m}
\!\times\! \widehat{x}}{x^4} \qquad (\vec{x} \neq 0) \; .
\end{equation}
The current rotates clockwise with respect to $\vec{m}$.

\section{de Sitter Geometry}

The de Sitter geometry is the unique, maximally symmetric solution
of Einstein's equation with a positive cosmological constant. Any
coordinatization $x^{\mu}$ of de Sitter can be described by a
mapping $x^{\mu} \to X^A(x)$ to the 4-dimensional submanifold of
5-dimensional Minkowski space such that,
\begin{equation}
-(X^0)^2 + (X^1)^2 + (X^2)^2 + (X^3)^2 + (X^4)^2 \equiv \eta_{AB}
X^A X^B = \frac1{H^2} \; .
\end{equation}
Here $H$ is the Hubble constant, which is related to the cosmological
constant by $\Lambda = 3 H^2$. The de Sitter metric in $x^{\mu}$
coordinates is given by,
\begin{equation}
g_{\mu\nu}(x) = \frac{\partial X^A}{\partial x^{\mu}}
\frac{\partial X^B}{\partial x^{\nu}} \eta_{AB} \; . \label{embed}
\end{equation}

The most convenient coordinates for our work are open conformal
coordinates, $x^{\mu} = (\eta,\vec{x})$. Whereas each of the spatial
coordinates $x^i$ runs from $-\infty$ to $+\infty$, the conformal
time runs from $\eta \rightarrow -\infty$ (the infinite past) to
$\eta \rightarrow 0$ (the infinite future). The 5-dimensional
embedding is,
\begin{eqnarray}
X^0 & = & \frac{a}{2 H} \Bigl[1 \!+\! H^2 ( x^2 \!-\! \eta^2) \Bigr]
\; , \label{confX0} \\
X^i & = & a x^i \; , \\
X^4 & = & \frac{a}{2 H} \Bigl[1 \!-\! H^2 ( x^2 \!-\! \eta^2) \Bigr]
\label{confX4} \; ,
\end{eqnarray}
where the scale factor is $a \equiv -1/H\eta$. The inverse
transformation is,
\begin{equation}
\eta = - \frac{\sqrt{\eta_{AB} X^A X^B}}{H (X^0 \!+\! X^4)} \qquad , \qquad
x^i = \frac{X^i}{H (X^0 \!+\! X^4)} \; . \label{inverse}
\end{equation}
From expression (\ref{embed}) we see that the invariant element is,
\begin{equation}
ds^2 = a^2(\eta) \Bigl[-d\eta^2 + d\vec{x} \!\cdot\! d\vec{x}\Bigr]
\qquad , \qquad a(\eta) = -\frac1{H \eta} \; .
\end{equation}
Hence $g_{\mu\nu} = a^2 \eta_{\mu\nu}$.

Because 4-dimensional electrodynamics is conformally invariant we can
reduce many aspects of the de Sitter computation (in conformal coordinates)
to familiar manipulations in flat space. One feature of conformal
coordinates that sometimes disturbs the mathematically inclined is that
they do not cover the full de Sitter manifold. One can see this by
adding (\ref{confX0}) and (\ref{confX4}) to get,
\begin{equation}
X^0 + X^4 = \frac{a}{H} > 0 \; .
\end{equation}
This is of no import because $\eta = {\rm const}$ defines a Cauchy
surface, so information from any part of the full de Sitter manifold
must enter the conformal coordinate submanifold as an initial
condition. The full de Sitter manifold is irrelevant if one regards
de Sitter as a paradigm for the geometry of primordial inflation.
From that perspective the conformal coordinate submanifold is just a
special case of the larger class of spatially flat,
Friedman-Robertson-Walker (FRW) geometries which are relevant to
inflationary cosmology.

We shall also be interested in static coordinates, $x^{\mu} =
(t,\vec{r})$. Although the time coordinates runs from $t \rightarrow
-\infty$ to $t \rightarrow +\infty$, the spatial radius obeys $0
\leq r < 1/H$. The 5-dimensional embedding is,
\begin{eqnarray}
X^0 & = & \frac1{H} \, \sinh(Ht) \sqrt{1 \!-\! H^2 r^2} \; ,
\label{stat1} \\
X^i & = & r^i \; , \label{stat2} \\
X^4 & = & \frac1{H} \, \cosh(Ht) \sqrt{1 \!-\! H^2 r^2} \; .
\label{stat3}
\end{eqnarray}
The inverse is,
\begin{equation}
t = \frac1{H} \, \tanh^{-1}\Bigl(\frac{X^0}{X^4}\Bigr) \qquad , \qquad
r^i = \frac{X^i}{H \sqrt{\eta_{AB} X^A X^B}} \; .
\end{equation}
Combining with (\ref{embed}) we find the invariant element,
\begin{equation}
ds^2 = -\Bigl[1 \!-\! H^2 r^2\Bigr] d\tau^2 + \frac{dr^2}{1 \!-\! H^2 r^2}
+ r^2 d\Omega^2 \; .
\end{equation}
Of course this is why the system is called ``static coordinates.''

As we have mentioned, mathematically oriented physicists sometimes
imagine that subtle errors must result from failing to formulate
quantum field theory on the full de Sitter manifold. Static
coordinates pose an equal peril for confusing those who seek to
deny the possibility of secular effects associated with the
continuous creation of massless, minimally coupled scalars and
gravitons during inflation. Quantum field theory cannot be
formulated in static coordinates because $t = {\rm const}$ does
not constitute a Cauchy surface. The static coordinate $t$ is the
proper time of an observer in free fall at the origin. Although no
point outside the static coordinate system can causally influence
this observer, that is only true for $\vec{r} = 0$. All other points
in the static coordinate system are in causal contact with points
from outside.

The preceding considerations preclude computing the vacuum polarization
in static coordinates, but they in no way prevent us from transforming
results from conformal coordinates into static coordinates. If the
source is at the static coordinate origin then an observer at fixed
$\vec{r}$ would appear, in conformal coordinates, to be moving
towards the origin with precisely the velocity needed to counteract
the expansion of spacetime and maintain a constant physical distance
from the source. Such an observer will experience effects associated
with boosting the conformal field strength, and other effects
associated with shrinking the conformal coordinate separation to
zero. By substituting (\ref{stat1}-\ref{stat3}) in (\ref{inverse}) we
see that the coordinate transformation is,
\begin{equation}
\eta = \frac{- e^{-Ht}}{H \sqrt{1 \!-\! H^2 r^2}} \qquad , \qquad
x^i = \frac{ r^i e^{-Ht}}{\sqrt{1 \!-\! H^2 r^2}} \; .
\end{equation}
The various components of the Jacobian are,
\begin{eqnarray}
\frac{\partial \eta}{\partial t} = \frac{e^{-Ht}}{\sqrt{1 \!-\! H^2 r^2}}
& , & \frac{\partial x^i}{\partial t} = -\frac{H r^i e^{-Ht}}{\sqrt{1
\!-\! H^2 r^2}} \; , \\
\frac{\partial \eta}{\partial r^j} = -\frac{H r^j e^{-Ht}}{[
1 \!-\! H^2 r^2]^{\frac32}} & , & \frac{\partial x^i}{\partial r^j}
= \frac{\delta^{ij} e^{-Ht}}{\sqrt{1 \!-\! H^2 r^2}} + \frac{H^2 r^i
r^j e^{-Ht}}{[1 \!-\! H^2 r^2]^{\frac32}} \; .
\end{eqnarray}
We denote the static coordinate field strengths with a tilde,
\begin{eqnarray}
\widetilde{F}_{0i} & = & \Bigl( \frac{e^{-2Ht}}{1 \!-\! H^2
r^2}\Bigr) \Biggl\{ F_{0i} \!+\! H r^j F_{ij} \Biggr\} \; ,
\label{staticE} \\
\widetilde{F}_{ij} & = & \Bigl( \frac{e^{-Ht}}{1 \!-\! H^2 r^2}\Bigr)^2
\Biggl\{ -H r^i F_{0j} \!+\! H r^j F_{0i} \!+\! (1 \!-\! H^2 r^2) F_{ij}
\nonumber \\
& & \hspace{6cm} + H^2 r^i r^k F_{kj} \!-\! H^2 r^j r^k F_{ki}
\Biggr\} . \qquad \label{staticB}
\end{eqnarray}
The factors of $e^{-2 Ht}$ seem to drive $\widetilde{F}_{\mu\nu}$ to
zero but one has to keep in mind that a factor of $1/x^2$ grows at
fixed $r$,
\begin{equation}
\frac1{x^2} = \frac{e^{2Ht}}{r^2} \Bigl[ 1 \!-\! H^2 r^2\Bigr] \; .
\end{equation}
Note also that the scale factor is,
\begin{equation}
a = e^{Ht} \sqrt{1 \!-\! H^2 r^2} \; .
\end{equation}

\section{Vacuum Polarization in de Sitter Space}

The plan of this section is the same as that of section 2. We begin
with some general considerations in which the one loop result for the
$[\mbox{}^{\mu} \Pi^{\nu}](x;x')$ is presented in open conformal
coordinates, and the one loop currents it engenders are evaluated
as much as possible for a general source. We then specialize the
classical source to a point charge and a point magnetic dipole. In
each case the one loop current density and the one loop field strengths
are first derived in open conformal coordinates, and then transformed
to static coordinates.

\subsection{General Considerations}

An important simplification associated with conformal coordinates
in $D=4$ spacetime dimensions is that all the scale factors in the
classical Maxwell equation cancel out,
\begin{equation}
\partial_{\nu} \Bigl[\sqrt{-g} \, g^{\nu\rho} g^{\mu\sigma}
F_{\rho\sigma} \Bigr] = \partial_{\nu} \Bigl[\eta^{\nu\rho}
\eta^{\mu\sigma} F_{\rho\sigma} \Bigr] \; .
\end{equation}
This means much of the flat space formalism of section 2 still applies.
In particular, the equation for the one loop correction to the field
strength is,
\begin{equation}
\partial^2 F^{(1)}_{\mu\nu} = \partial_{\mu} J^{(1)}_{\nu} -
\partial_{\nu} J^{(1)}_{\mu} \; , \label{dSeqn}
\end{equation}
where $\partial^2 \equiv \eta^{\mu\nu} \partial_{\mu} \partial_{\nu} =
-\partial_{\eta}^2 + \nabla^2$ is the flat space d'Alembertian and
the one loop current is,
\begin{equation}
J^{(1)}_{\mu}(x) \equiv \eta_{\mu\rho} \int \!\! d^4x' \, \Bigl[
\mbox{}^{\rho} \Pi^{\nu}_{(1)} \Bigr](x;x') A^{(0)}_{\nu}(x') \; .
\end{equation}
The retarded solution for (\ref{dSeqn}) is,
\begin{equation}
F^{(1)}_{\mu\nu}(x) = - \int \!\! d^4x' \,
\frac{\delta(\Delta \eta \!-\! \Delta x)}{4 \pi \Delta x} \Bigl[
\partial'_{\mu} J^{(1)}_{\nu}(x') \!-\! \partial'_{\nu}
J^{(1)}_{\mu}(x') \Bigr] \; , \label{retsol}
\end{equation}
where the conformal coordinate separations are $\Delta \eta \equiv \eta
- \eta'$ and $\Delta x \equiv \Vert \vec{x} - \vec{x}' \Vert$.

Because the massless, minimally coupled scalar is not de Sitter
invariant, the vacuum polarization it engenders contains two distinct
tensor structures. One of these is proportional to the covariant
transverse projection operator, whereas the other one is proportional
to the purely spatial transverse projection operator constructed
from $\overline{\eta}^{\mu\nu} \equiv \eta^{\mu\nu} + \delta^{\mu}_0
\delta^{\nu}_0$,\footnote{One might wonder if there is any advantage
to employing a covariant representation based on covariant derivatives
and de Sitter invariant basis tensors. However, a systematic
investigation of this formalism reveals that it is cumbersome and
that it obscures the essential physics \cite{LPW1}. The procedure
for converting our noncovariant --- but simple --- representation
(\ref{noncovPi}) to the complicated and counter-intuitive covariant
representation can be found in \cite{LPW2}.}
\begin{eqnarray}
\lefteqn{\Bigl[\mbox{}^{\mu} \Pi^{\nu}\Bigr](x;x') =
\Bigl[ \eta^{\mu\nu} \eta^{\rho\sigma} \!\!-\! \eta^{\mu\rho}
\eta^{\nu\sigma} \Bigr] \partial'_{\rho} \partial_{\sigma}
\Biggl\{ \Pi_F(x;x') \!+\! \Pi_C(x;x') \!+\! \Pi_G(x;x')\Biggr\} }
\nonumber \\
& & \hspace{6cm} + \Bigl[ \overline{\eta}^{\mu\nu}
\overline{\eta}^{\rho\sigma} \!-\! \overline{\eta}^{\mu\rho}
\overline{\eta}^{\nu\sigma} \Bigr] \partial'_{\rho} \partial_{\sigma}
\Pi_K(x;x') \; . \qquad \label{noncovPi}
\end{eqnarray}
It is convenient to break the covariant structure function up into
the old, flat space contribution $\Pi_F$, a term $\Pi_C$ like the
conformal anomaly, and a nonlocal de Sitter contribution $\Pi_G$.
At one loop order one has \cite{PTW,PW2},
\begin{eqnarray}
\Pi^{(1)}_F(x;x') & = & -\frac{\alpha}{96 \pi^2} \, \partial^4
\Biggl[ \theta( \Delta \eta \!-\! \Delta x) \Biggl\{ \ln\Bigl[\mu^2
( \Delta \eta^2 \!-\! \Delta x^2)\Bigr] \!-\! 1\Biggr\} \Biggr] \; ,
\qquad \label{PiF} \\
\Pi^{(1)}_C(x;x') & = & -\frac{\alpha}{6 \pi} \, \ln(a) \delta^4(x \!-\! x')
\; , \label{PiC} \\
\Pi^{(1)}_G(x;x') & = & -\frac{\alpha H^2 a}{8 \pi^2} \, \partial^2
\Biggl[ a' \theta(\Delta \eta \!-\! \Delta x) \Biggl\{ \ln\Bigl[
H^{\prime 2} (\Delta \eta^2 \!-\! \Delta x^2)\Bigr] \!+\! 1\Biggr\}
\Biggr] \; . \qquad \label{PiG}
\end{eqnarray}
The one loop noncovariant structure function is \cite{PTW,PW2},
\begin{equation}
\Pi^{(1)}_K(x;x') = \frac{\alpha H^4 a^2 a^{\prime 2}}{4 \pi^2} \,
\theta(\Delta \eta \!-\! \Delta x) \Biggl\{ \ln\Bigl[ H^{\prime 2}
(\Delta \eta^2 \!-\! \Delta x^2)\Bigr] \!+\! 2 \Biggr\} \; .
\label{PiK}
\end{equation}
Here $H' \equiv e^{\gamma - 1} H$, where $\gamma \approx 0.57721$ is
Euler's constant. Although the spatial dependence of the various structure
functions is limited to the coordinate separation, $\vec{x} - \vec{x}'$,
factors of $a$ and $a'$ complicate the temporal dependence.

The various one loop current densities can be expressed in terms of
the classical current density $J_{\mu}$ and the classical field strength
$F^{(0)}_{\mu\nu}$,
\begin{eqnarray}
J^{1F}_{\mu}(x) & = & -\frac{\alpha \partial^4}{96 \pi^2} \!\int \!\! d^4x' \,
\theta(\Delta \eta \!-\! \Delta x) \Biggl\{ \ln\Bigl[\mu^2 (\Delta \eta^2
\!-\! \Delta x^2)\Bigr] \!-\! 1\Biggr\} J_{\mu}(x') \; , \qquad \\
J^{1C}_{\mu}(x) & = & -\frac{\alpha}{6 \pi} \partial^{\nu} \Bigl[
\ln(a) F^{(0)}_{\nu\mu}(x) \Bigr] = \frac{\alpha}{6\pi} \Bigl[-
\ln(a) J_{\mu}(x) \!+\! H a F^{(0)}_{0 \mu}(x) \Bigr] \; , \qquad
\label{J1C} \\
J^{1G}_{\mu}(x) & = & -\frac{\alpha H^2}{8 \pi^2} \partial^{\nu} \Biggl[
a \partial^2 \!\! \int \!\! d^4x' \, a' \theta(\Delta \eta \!-\!
\Delta x) \nonumber \\
& & \hspace{3.5cm} \times \Biggl\{ \ln\Bigl[ H^{\prime 2} (\Delta \eta^2
\!-\! \Delta x^2) \Bigr] \!+\! 1 \Biggr\} F^{(0)}_{\nu\mu}(x') \Biggr]
\; , \qquad \\
J^{1K}_{\mu}(x) & = & \frac{\alpha H^4 a^2}{4 \pi^2} \!\! \int \!\! d^4x' \,
a^{\prime 2} \theta(\Delta \eta \!-\! \Delta x) \nonumber \\
& & \hspace{3.5cm} \times \Biggl\{ \ln\Bigl[H^{\prime 2} (\Delta
\eta^2 \!-\! \Delta x^2)\Bigr] \!+\! 2\Biggr\} \partial_j'
F^{(0)}_{j \bar{\mu}}(x') \; . \qquad \label{J1K}
\end{eqnarray}
Of course $J^{1F}_{\mu}(x)$ is the same as the flat space result of section 2.
Note that $J^{1K}_{0}(x) = 0$ because the index $\bar{\mu}$ is purely spatial.
It is useful to $3+1$ decompose $J^{1G}_{\mu}(x)$,
\begin{eqnarray}
J^{1G}_0(x) & \!\!\!\!=\!\!\!\! & -\frac{\alpha H^2 a}{8 \pi^2} \partial^2
\!\!\! \int \!\! d^4x' \, a' \theta(\Delta \eta \!-\! \Delta x) \Biggl\{\!
\ln\Bigl[ H^{\prime 2} (\Delta \eta^2 \!-\! \Delta x^2)\Bigr] \!+\! 1 \!
\Biggr\} J_0(x') \; , \quad \label{J1G0} \\
J^{1G}_i(x) & \!\!\!\!=\!\!\!\! & -\frac{\alpha H^2 a}{8 \pi^2} \partial^2
\!\!\! \int \!\! d^4x' \, a' \theta(\Delta \eta \!-\! \Delta x) \Biggl\{ \!
\ln\Bigl[ H^{\prime 2} (\Delta \eta^2 \!-\! \Delta x^2)\Bigr] \!+\! 1 \!
\Biggr\} J_i(x') \; , \quad \nonumber \\
& & \hspace{-.8cm} +\frac{\alpha H^3 a^2}{8 \pi^2} \partial^2 \!\! \int \!\!
d^4x' \, a' \theta(\Delta \eta \!-\! \Delta x) \Biggl\{ \ln\Bigl[ H^{\prime 2}
(\Delta \eta^2 \!-\! \Delta x^2)\Bigr] \!+\! 1\Biggr\} F^{(0)}_{0i}(x')
\; , \qquad \nonumber \\
& & \hspace{-.8cm} +\frac{\alpha H^3 a}{8 \pi^2} \partial^2 \!\!
\int \!\! d^4x' \, a^{\prime 2} \theta(\Delta \eta \!-\! \Delta x)
\Biggl\{ \ln\Bigl[ H^{\prime 2} (\Delta \eta^2 \!-\! \Delta
x^2)\Bigr] \!+\! 1\Biggr\} F^{(0)}_{0i}(x') \; . \qquad \label{J1Gi}
\end{eqnarray}

We should also comment on the range of coordinates. Unlike the flat
space case, it is not possible to release this system in free vacuum
infinitely far back in the past because perturbation theory breaks
down after a finite interval of conformal time \cite{PTsW}. The one
loop results (\ref{PiF}-\ref{PiK}) were computed with initial time
$\eta_I = -1/H$, corresponding to $a_I = 1$. For a source at the
origin, the factors of $\theta(\Delta \eta - \Delta x)$ in
(\ref{PiF}-\ref{PiK}) imply that there is no one loop effect for
radii $x$ outside the range,
\begin{equation}
\eta_I < \eta \!-\! x \qquad \Longrightarrow \qquad H x < 1 \!-\!
\frac1{a} \; .
\end{equation}
For future reference we define the scale factors $a_{\pm}$ evaluated
at conformal times $\eta \pm x$,
\begin{equation}
a_{\pm} \equiv -\frac1{H (\eta \!\pm\! x)} = \frac1{a^{-1} \!\mp\!
Hx} = e^{H t} \sqrt{ \frac{1 \!\pm\! Hr}{1 \!\mp\! Hr}} \;.
\label{apm}
\end{equation}
The late time behavior of these scale factors depends critically
upon whether or not one fixes the conformal radius $x$ or static
radius $r$,
\begin{equation}
\Bigl( a_{\pm} \Bigr)_{x\ {\rm fixed}} \longrightarrow \mp \frac1{H
x} \qquad , \qquad \Bigl( a_{\pm} \Bigr)_{r\ {\rm fixed}}
\longrightarrow e^{Ht} \sqrt{ \frac{1 \!\pm\! Hr}{1 \!\mp\! Hr}} \;
.
\end{equation}

\subsection{Response to a point charge}\label{pointQ}

The electromagnetic coupling to a point charge $q$ with worldline
$\chi^{\mu}(\tau)$ is,
\begin{equation}
S_q = q \int \!\! d\tau \, \dot{\chi}^{\mu} A_{\mu}(\chi) \; .
\end{equation}
Because $S_q$ is independent of the metric, the current density of a
point charge is unchanged from (\ref{charge}), nor are the 0th order
field strengths changed from section 2.1,
\begin{equation}
J^{\mu}(\eta,\vec{x}) = q \delta^{\mu}_0 \delta^3(\vec{x}) \quad ,
\quad F^{(0)}_{0i}(\eta,\vec{x}) = \frac{q}{4\pi} \frac{x^i}{x^3}
\quad , \quad F^{(0)}_{ij}(\eta,\vec{x}) = 0 \; . \label{charge0}
\end{equation}
One consequence is that $J^{1F}_{\mu}$ must agree with expression
(\ref{1loopJ}), and the associated one loop field strengths must
agree with (\ref{Efield}-\ref{nofield}),
\begin{equation}
J^{1F}_{\mu}(\eta,\vec{x}) = \frac{\alpha q}{24 \pi^2}
\frac{\delta^0_{\mu}}{x^3} \; (\vec{x} \neq 0) \; , \;
F^{1F}_{\mu\nu}(\eta,\vec{x}) = -\frac{\alpha}{6\pi} \Bigl[ \ln(2
\mu x) \!-\! 1\Bigr] \times F^{(0)}_{\mu\nu}(\eta,\vec{x}) \; .
\label{E1F}
\end{equation}

Substituting (\ref{charge0}) into (\ref{J1C}) gives,
\begin{equation}
J^{1C}_{0}(\eta,\vec{x}) = \frac{\alpha q}{6\pi} \, \ln(a)
\delta^3(\vec{x}) \quad , \quad J^{1C}_{i}(\eta,\vec{x}) =
\frac{\alpha q H a}{24 \pi^2} \frac{x^i}{x^3} = -\partial_i \Biggl\{
\frac{\alpha q}{24 \pi^2} \frac{Ha}{x} \Biggr\} .
\end{equation}
Hence the $C$-type charge density represents a fractional screening
of the classical charge density by $-\frac{\alpha}{6\pi} \ln(a)$,
with the current density $J^{1C}_i$ carrying off the positive charge
to infinity. We can easily solve for the $C$-type electric field by
noting,
\begin{equation}
\partial^2 F^{1C}_{0i}(\eta,\vec{x}) = \partial_0
J^{1C}_i(\eta,\vec{x}) \!-\! \partial_i J^{1C}_0(\eta,\vec{x}) =
\partial^2 \partial_i \Biggl\{ \frac{\alpha q}{24 \pi^2}
\frac{\ln(a)}{x} \Biggr\} .
\end{equation}
Therefore, the $C$-type field strengths are,
\begin{equation}
F^{1C}_{\mu\nu}(\eta,\vec{x}) = -\frac{\alpha}{6\pi} \ln(a) \times
F^{(0)}_{\mu\nu}(\eta,\vec{x}) \; . \label{E1C}
\end{equation}

Substituting (\ref{charge0}) into (\ref{J1K}) implies that all
components of the $K$-type current density vanish. It is better to
approach the $G$-type current density and its associated field
strength indirectly. Suppose the temporal and spatial components of
the current density take the form,
\begin{equation}
J^{1G}_0(\eta,\vec{x}) = f(\eta,x) \qquad , \qquad
J^{1G}_i(\eta,\vec{x}) = \partial_i g(\eta,x) \; .
\end{equation}
Current conservation implies,
\begin{equation}
0 = -\partial_{\eta} f(\eta,x) + \nabla^2 g(\eta,x) \;
\Longrightarrow \; g(\eta,x) = \int_0^{x} \!\! dx' \Bigl(1 \!-\!
\frac{x'}{x}\Bigr) x'
\partial_{\eta} f(\eta,x') \; .
\end{equation}
The magnetic field is obviously zero and we can find the electric
field by noting,
\begin{eqnarray}
\partial^2 F^{1G}_{0i}(\eta,\vec{x}) & = & \partial_{\eta}
J^{1G}_i(\eta,\vec{x}) \!-\! \partial_i J^{1G}_0(\eta,\vec{x}) \; ,
\\
& = & \partial_i \Biggl\{ \partial_{\eta}^2 \frac1{\nabla^2} f \!-\!
f\Biggr\} \; , \\
& = & -\partial^2 \partial_i \int_0^{x} \!\! dx' \Bigl(1 \!-\!
\frac{x'}{x}\Bigr) x' f(\eta,x') \; .
\end{eqnarray}
Hence everything follows from the zero component,
\begin{eqnarray}
J^{1G}_i(\eta,\vec{x}) & = & \frac{x^i}{x^3} \int_0^{x} \!\! dx'
{x'}^2 \partial_{\eta} J^{1G}_0(\eta,x') \; , \label{Jvec1G} \\
F^{1G}_{0i}(\eta,\vec{x}) & = & -\frac{x^i}{x^3} \int_0^{x} \!\! dx'
{x'}^2 J^{1G}_0(\eta,x') \; , \label{E1G} \\
F^{1G}_{ij}(\eta,\vec{x}) & = & 0 \; . \label{B1G}
\end{eqnarray}
Expression (\ref{E1G}) is recognizable as the integral form of
Gauss's law, and expression (\ref{Jvec1G}) is the Maxwell
displacement current.

It remains to evaluate $J^{1G}_0(\eta,x)$,
\begin{equation}
J^{1G}_0(\eta,x) = \frac{\alpha q H^2 a}{8 \pi^2} \partial^2
\!\int_{\eta_I}^{\eta - x} \!\!\!\!\! d\eta' a' \Biggl\{ \ln\Bigl[H'
(\Delta \eta \!-\! x)\Bigr] \!+\! \ln\Bigl[ H' (\Delta \eta \!+\!
x)\Bigr] \!+\! 1\Biggr\} . \label{J1G01}
\end{equation}
Expression (\ref{J1G01}) is sufficiently intricate that the best
strategy is to treat each of the three integrands separately, and to
make an additional distinction between the two derivative operators
which result when the d'Alembertian is specialized to functions of
just $\eta$ and $x$ as in (\ref{DAL}),
\begin{equation}
\partial^2 \longrightarrow \frac1{2x} (\partial_{\eta} \!-\!
\partial_x) (\partial_{\eta} \!+\! \partial_x) (\eta \!-\! x)
- \frac1{2x} (\partial_{\eta} \!-\! \partial_x) (\partial_{\eta}
\!+\! \partial_x) (\eta \!+\! x) \; . \label{newDAL}
\end{equation}
The first of the integrands in (\ref{J1G01}) depends only upon $\eta
- x$ so it vanishes when acted upon by the first operator of
(\ref{newDAL}),
\begin{equation}
\frac1{2x} (\partial_{\eta} \!-\! \partial_x) (\partial_{\eta} \!+\!
\partial_x) (\eta \!-\! x) \int_{\eta_I}^{\eta-x} \!\!\!\!\! d\eta' a'
\ln\Bigl[H' (\Delta \eta \!-\! x)\Bigr] = 0 \; . \label{1A}
\end{equation}
A nonzero result emerges when this same first integrand is acted on
by the second operator in (\ref{newDAL}),
\begin{eqnarray}
\lefteqn{ -\frac1{2x} (\partial_{\eta} \!-\! \partial_x)
(\partial_{\eta} \!+\! \partial_x) (\eta \!+\! x)
\int_{\eta_I}^{\eta-x} \!\!\!\!\! d\eta' a' \ln\Bigl[H' (\Delta \eta
\!-\! x)\Bigr] \nonumber} \\
& & \hspace{5cm} = -\frac1{x} (\partial_{\eta} \!-\! \partial_x)
\int_{\eta_I}^{\eta - x} \!\!\!\!\! d\eta' a' \ln\Bigl[H' (\Delta
\eta \!-\! x)\Bigr] \; , \qquad \\
& & \hspace{5cm} = -\frac{2 a_{-}}{x} \Biggl\{ \gamma \!-\! 1 \!-\!
\ln(a_{-}) \!+\! \ln\Bigl(1 \!-\! \frac1{a_{-}}\Bigr) \Biggr\} .
\qquad \label{1B}
\end{eqnarray}
(We remind the reader of the scale factors $a_{\pm}$ defined in
(\ref{apm}).) Acting the first part of the d'Alembertian
(\ref{newDAL}) on the second integrand in (\ref{J1G01}) gives,
\begin{eqnarray}
\lefteqn{\frac1{2x} (\partial_{\eta} \!-\! \partial_{x})
(\partial_{\eta} \!+\! \partial_{x}) (\eta \!-\! x)
\int_{\eta_I}^{\eta - x} \!\!\!\!\! d\eta' a' \ln\Bigl[ H' (\Delta
\eta \!+\! x)\Bigr] \nonumber } \\
& & \hspace{2cm} = \frac1{x} (\partial_{\eta} \!-\! \partial_{x})
(\eta \!-\! x) \int_{\eta_I}^{\eta - x} \!\!\!\!\! d\eta' a'
\frac1{\Delta \eta \!+\! x} , \qquad \\
& & \hspace{2cm} = \frac{a_{+}}{x} (\partial_{\eta} \!-\!
\partial_{x}) (\eta \!-\! x) \Biggl\{-\ln(a_{-}) \!-\! \ln(2 H x)
\!+\! \ln\Bigl(1 \!-\! \frac1{a_{+}}\Bigr) \Biggr\} \; , \qquad \\
& & \hspace{2cm} = \frac{2 a_{+}}{x} \Biggl\{-\ln(a_{-}) \!-\! \ln(2
H x) \!+\! \ln\Bigl(1 \!-\! \frac1{a_{+}}\Bigr) \!+\! \frac12 \!-\!
\frac1{2 a H x} \Biggr\} . \qquad \label{2A}
\end{eqnarray}
The second part of the d'Alembertian (\ref{newDAL}) acts on this
second integrand to produce,
\begin{eqnarray}
\lefteqn{-\frac1{2x} (\partial_{\eta} \!-\! \partial_{x})
(\partial_{\eta} \!+\! \partial_{x}) (\eta \!+\! x)
\int_{\eta_I}^{\eta - x} \!\!\!\!\! d\eta' a' \ln\Bigl[ H' (\Delta
\eta \!+\! x)\Bigr] \nonumber } \\
& & \hspace{6cm} = -\frac{a_{-}}{x} (\partial_{\eta} \!+\!
\partial_{x}) (\eta \!+\! x) \ln(2 H x) \; , \qquad \\
& & \hspace{6cm} = -\frac{2 a_{-}}{x} \Biggl\{ \ln(2 H x) \!+\!
\frac12 \!-\! \frac1{2 a H x} \Biggr\} . \qquad \label{2B}
\end{eqnarray}
And the final term of the integrand in (\ref{J1G01}) is simple
enough that the d'Alem\-bert\-ian can be kept together,
\begin{equation}
\partial^2 \int_{\eta_I}^{\eta - x} \!\!\!\!\! d\eta' a' =
\partial^2 \Biggl\{ \frac{\ln(a_{-})}{H} \Biggr\} = -\frac{2 a_{-}}{x}
\; . \label{3AB}
\end{equation}

Summing expressions (\ref{1A}), (\ref{1B}), (\ref{2A}), (\ref{2B})
and (\ref{3AB}), and multiplying by the prefactor of (\ref{J1G01})
gives the $G$-type charge density,
\begin{eqnarray}
\lefteqn{J^{1G}_{0}(\eta,x) = \frac{\alpha q H^2 a}{4 \pi^2}
\Biggl\{ \frac{a_{-}}{x} \Biggl[ \ln(a_{-}) \!-\! \ln(2 H x) \!-\!
\ln\Bigl[1 \!-\! \frac1{a_{-}}\Bigr] \!-\! \gamma \!-\! \frac12
\!+\! \frac1{2 a H x}\Biggr] } \nonumber \\
& & \hspace{2.5cm} + \frac{a_{+}}{x} \Biggl[ -\ln(a_{-}) \!-\! \ln(2
H x) \!+\! \ln\Bigl[1 \!-\! \frac1{a_{+}}\Bigr] \!+\! \frac12 \!-\!
\frac1{2 a H x}\Biggr] \Biggr\} . \qquad \label{J1G02}
\end{eqnarray}
Expression (\ref{J1G02}) is complicated but its import becomes clear
at late times, either at fixed $x$ or at fixed $r = x/a$,
\begin{eqnarray}
\Bigl( J^{1G}_0 \Bigr)_{\eta \gg \eta_I \atop x \; {\rm fixed}} &
\!\!=\!\! & \frac{\alpha q H a}{4 \pi^2 x^2} \Biggl\{ 2 \ln\Bigl(
\frac1{H x}\Bigr) \!-\! \ln\Bigl(1 \!-\! H^2 x^2\Bigr) \!+\! O(1)
\Biggr\} ,
\qquad \\
\Bigl( J^{1G}_0 \Bigr)_{\eta \gg \eta_I \atop x = r/a} & \!\!=\!\! &
\frac{\alpha q H^2 a^3}{4 \pi^2 r} \Biggl\{ \frac{2 \ln(a)}{1 \!+\!
Hr} \!+\! \frac{2 H r \ln(1 \!+\! Hr)}{1 \!-\! H^2 r^2} \!-\!
\frac{2 \ln(2 H r)}{1 \!-\! H^2 r^2} \!+\! O(1) \Biggr\} . \qquad
\end{eqnarray}
In each case we see that the one loop charge density (which is {\it
minus} $J^{1G}_0$) is opposite to $q$, implying screening. We also
see that the effect from $J^{1G}_0$ is much stronger than that of
$J^{1C}_0$ (powers of $a$ versus $\ln(a)$), which is itself stronger
than that of $J^{1F}_0$.

The $G$-type contribution to the one loop electric field follows
from substituting each of the ten terms of expression (\ref{J1G02})
in (\ref{E1G}). The integrals are not illuminating and have been
consigned to expressions (\ref{Int1}-\ref{Int10}) of Appendix A.
After considerable rearrangement the final result is,
\begin{eqnarray}
\lefteqn{ \hspace{-.3cm} F^{1G}_{0i}(\eta,\vec{x}) =
-\frac{\alpha}{\pi} \times F^{(0)}_{0i}(\eta,\vec{x}) \times
\Biggl\{ a H x \Biggl[ (1 \!-\! \gamma) \Bigl[1 \!-\!
\ln\Bigl(\frac1{a} \!+\! H x\Bigr)\Bigr] }
\nonumber \\
& & \hspace{3cm} + 2 - \Bigl( \frac{a \!-\! 1}{a H x}\Bigr)
\ln\Biggl( \frac{1 \!+\! \frac{a H x}{a \!-\! 1}}{1 \!-\! \frac{a H
x}{a \!-\! 1}} \Biggr) - \ln\Bigl[ \Bigl(1 \!-\! \frac1{a}\Bigr)^2
\!-\! H^2 x^2\Bigr] \Biggr] \nonumber \\
& & \hspace{-.5cm} - \ln\Bigl[1 \!+\! a H x\Bigr] \Biggl[2 \ln(a)
\!-\! \frac12 \ln\Bigl[1 \!+\! a H x\Bigr] \!+\! 1 \!-\! \gamma
\!-\! \ln(2) \Biggr] -\frac12 {\rm Li}_2\Bigl[1 \!-\! a^2 H^2
x^2\Bigr] \nonumber \\
& & \hspace{.7cm} + {\rm Li}_2\Bigl[\frac12 \!-\! \frac12 a H
x\Bigr] \!-\! \frac{\pi^2}{12} \!+\! \frac12 \ln^2(2) \!+\! {\rm
Li}_2\Bigl[ \frac1{a} \!-\! H x\Bigr] \!-\! {\rm Li}_2\Bigl[
\frac1{a} \!+\! H x\Bigr] \Biggr\} . \label{fullE1G}
\end{eqnarray}
The symbol ${\rm Li}_2(z)$ denotes the dilogarithm function,
\begin{equation}
{\rm Li}_2(z) \equiv \int_{z}^0 \!\! dt \frac{\ln(1 \!-\! t)}{t} =
\sum_{k=1}^{\infty} \frac{z^k}{k^2} = -\frac12 \ln^2(-z) \!-\!
\frac{\pi^2}{6} \!-\! \sum_{k=1}^{\infty} \frac1{k^2 z^k} \; .
\label{dilog}
\end{equation}
Note that we have employed the identity,
\begin{equation}
{\rm Li}_2(-x) - {\rm Li}_2(1 \!-\! x) = -\ln(x) \ln(1 \!+\! x)
-\frac{\pi^2}{6} - \frac12 {\rm Li}_2(1 \!-\! x^2) \; .
\end{equation}

Expression (\ref{fullE1G}) is unwieldy. One can understand it better
by taking the limit of late time at fixed $x$,
\begin{equation}
\Bigl( F^{1G}_{0i} \Bigr)_{x \, {\rm fixed}} \longrightarrow
-\frac{\alpha}{\pi} \times \frac{q}{4 \pi} \frac{x^i}{x^3} \times a
H x \times \mathcal{F}(Hx) \; . \label{xfix}
\end{equation}
The proportionality function is positive definite,
\begin{equation}
\mathcal{F}(Hx) = (1 \!-\! \gamma) \Bigl[1 \!-\! \ln(Hx) \Bigr] + 2
- \frac1{Hx} \ln\Bigl[ \frac{1 \!+\! Hx}{1 \!-\! Hx}\Bigr] - \ln(1
\!-\! H^2 x^2) \; . \label{Fdef}
\end{equation}
(Recall that $\gamma \approx 0.57721$ is Euler's constant, so $1 -
\gamma \approx 0.42278$.) For small $Hx$ it has the expansion,
\begin{equation}
\mathcal{F}(Hx) = (1 \!-\! \gamma) \ln\Bigl[\frac1{H x}\Bigr] + (1
\!-\! \gamma) + \frac43 (Hx)^2 + \frac{3}{10} (H x)^4 + \dots
\end{equation}
The largest $Hx$ can get is one, at which point,
\begin{equation}
\mathcal{F}(1) = 3 \!-\! \gamma \!-\! 2 \ln(2) \approx 1.03649 \; .
\end{equation}

Combining expressions (\ref{E1F}), (\ref{E1C}) and (\ref{xfix})
gives the total one loop correction to the electric field at late
times for fixed co-moving position $x$,
\begin{equation}
\Bigl( F^{(1)}_{0i} \Bigr)_{x \, {\rm fixed}} \longrightarrow
-\frac{\alpha}{\pi} \times F^{(0)}_{0i} \Biggl\{ \frac16 \Bigl[
\ln(2 \mu x) \!-\! 1\Bigr] + \frac16 \ln(a) + a H x \times
\mathcal{F}(Hx) \Biggr\} . \label{E1late}
\end{equation}
It will be seen that the $1G$ contribution totally dominates the
$1C$ and $1F$ contributions. The one loop correction we have found
is consistent with screening from the vacuum polarization induced by
the vast ensemble of charged scalars produced during inflation. The
presence of a scale factor means that the one loop correction
cancels the tree order effect after only about $\ln(a) \sim
\ln(\pi/\alpha) \approx 6$ e-foldings! Of course this does not mean
fixed $x$ observers see the field of a negative charge after that
time; what happens instead is that they see effectively zero field
strength. Perturbation theory breaks down because screening has
become nonperturbatively strong.

The result is curiously different at fixed $r$ in static
coordinates. To see why, let us first evaluate the $1G$ contribution
(\ref{fullE1G}) at late times holding $r = a H x$ fixed,
\begin{equation}
\Bigl( F^{1G}_{0i} \Bigr)_{r \, {\rm fixed}} \longrightarrow
-\frac{\alpha}{\pi} \times \frac{q}{4\pi} \frac{a^2 r^i}{r^3} \times
\ln(a) \Bigl[- 2\ln(1 \!+\! Hr) + (1 \!-\! \gamma) Hr\Bigr] \; .
\label{rfix}
\end{equation}
This is negative definite for all $0 < Hr < 1$, which is consistent
with anti-screening. Of course that is what happens in flat space as
well, when one gets very close to the source. Note that the factor
of $\ln(a)$ cancels between the $1F$ and $1C$ contributions at fixed
$r$,
\begin{equation}
\frac16 \Bigl[ \ln(2 \mu x) \!-\! 1\Bigr] + \frac16 \ln(a) = \frac16
\Bigl[ \ln(2 \mu r) \!-\! 1\Bigr] \; .
\end{equation}
Hence the $1G$ contribution is still dominant.

To derive the electric field strength in static coordinates we must
still transform the vector indices. Substituting (\ref{rfix}) into
expression (\ref{staticE}) gives,
\begin{eqnarray}
\Bigl( \widetilde{F}^{(1)}_{0i} \Bigr)_{r \, {\rm fixed}} &
\longrightarrow & -\frac{\alpha}{\pi} \times \frac{q}{4\pi} \frac{
r^i}{r^3} \times H t \Bigl[- 2\ln(1 \!+\! Hr) + (1 \!-\! \gamma)
Hr\Bigr] \; , \\
& = & \widetilde{F}^{(0)}_{0i} \times \frac{\alpha}{\pi} \times H t
\Bigl[2\ln(1 \!+\! Hr) - (1 \!-\! \gamma) Hr\Bigr] \; .
\label{E1tilde}
\end{eqnarray}
We see that the one loop correction at fixed static coordinate $r$
(\ref{E1tilde}) has both the opposite sign and a much slower growth
than the result in conformal coordinates (\ref{E1late}). The
physical interpretation seems to be that the static coordinate
observer experiences the logarithmic running of the electromagnetic
coupling which is built into the Bunch-Davies vacuum, no matter what
coordinate system one employs. The effect becomes nonperturbatively
strong after about $H t \sim \pi/\alpha \approx 430$ e-foldings.

\subsection{Response to a point magnetic dipole}\label{pointM}

In conformal coordinates the current density and classical field
strengths of a point dipole are unchanged from flat space,
\begin{eqnarray}
J^0(\eta,\vec{x}) = 0 & , & \vec{J}(\eta,\vec{x}) = -\vec{m} \times
\vec{\nabla} \delta^3(\vec{x}) \; , \label{magdipole} \\
F^{(0)}_{0i}(\eta,\vec{x}) = 0 & , & F^{(0)}_{ij}(\eta,\vec{x}) =
\frac{\epsilon^{ijk}}{4\pi} \Bigl[m^k \nabla^2 \!-\! \vec{m}
\!\cdot\! \vec{\nabla} \partial_k \Bigr] \Biggl\{ -\frac1{x}
\Biggr\} \; . \label{magfield}
\end{eqnarray}
It follows that the $1F$ contributions to the one loop induced
current density and field strengths are the same as we found in
section 2.3,
\begin{eqnarray}
J^{1F}_0(\eta,\vec{x}) = 0 & , & \vec{J}^{1F}(\eta,\vec{x}) =
-\frac{\alpha}{8 \pi^2} \frac{\vec{m} \! \times \! \widehat{x}}{x^4}
\; , \qquad \label{J1F} \\
F^{1F}_{0i}(\eta,\vec{x}) = 0 & , & F^{1F}_{ij}(\eta,\vec{x}) =
\frac{\alpha \epsilon^{ijk}}{24 \pi^2} \Bigl[m^k \nabla^2 \!-\!
\vec{m} \!\cdot\! \vec{\nabla} \partial_k \Bigr] \Biggl\{ - \frac{
\ln(2 \mu x)}{x} \Biggr\} . \qquad \label{BM1F}
\end{eqnarray}
Note that we have neglected delta function terms in (\ref{J1F}).

Substituting (\ref{magdipole}-\ref{magfield}) in expression
(\ref{J1C}) gives the $1C$ contribution to the one loop induced
current density,
\begin{equation}
J^{1C}_0(\eta,\vec{x}) = 0 \qquad , \qquad J^{1C}_i(\eta,\vec{x}) =
\frac{\alpha \ln(a)}{6 \pi} \, \epsilon^{ijk} m^j \partial_k
\delta^3(\vec{x}) \; .
\end{equation}
Using these currents, with a few partial integrations, in relation
(\ref{retsol}) produces the associated field strengths,
\begin{eqnarray}
F^{1C}_{0i}(\eta,\vec{x}) & = & \frac{\alpha \epsilon^{ijk}}{24 \pi^2}
\, m^j \partial_k \partial_0 \Biggl\{ \frac{ \ln(Hx \!+\! \frac1{a})}{x}
\Biggr\} \; , \label{EM1C} \\
F^{1C}_{ij}(\eta,\vec{x}) & = & \frac{\alpha \epsilon^{ijk}}{24 \pi^2}
\Bigl[m^k \nabla^2 \!-\! \vec{m} \!\cdot\!  \vec{\nabla} \partial_k \Bigr]
\Biggl\{ \frac{\ln(Hx \!+\! \frac1{a})}{x} \Biggr\} \; . \qquad \label{BM1C}
\end{eqnarray}
Recall that we are assuming $0 < H x < 1 - 1/a$.

It is best to add the $G$ and $K$ currents together before inferring
the field strengths they induce, so we derive them together. From
expression (\ref{J1G0}) we see that $J^{1G}_0(\eta,\vec{x})$ vanishes
because the classical charge density is zero for a point magnetic
dipole. Substituting the other classical values
(\ref{magdipole}-\ref{magfield}) into expression (\ref{J1Gi}) and
performing some partial integrations reduces the induced $G$-type
current density to a single integral over conformal time,
\begin{equation}
J^{1G}_i(\eta,\vec{x}) = \frac{\alpha H^2 a}{8 \pi^2}
\partial^2 \epsilon^{ijk} m^j \partial_k \int_{\eta_I}^{\eta - x}
\!\! d\eta' \, a' \Biggl\{ \ln\Bigl[ H^{\prime 2} (\Delta \eta^2
\!-\! \Delta x^2)\Bigr] \!+\! 1 \Biggr\} \; . \label{J1BG0}
\end{equation}
Of course the temporal component of $J^{1K}_{\mu}(\eta,\vec{x})$
always vanishes. From expression (\ref{J1K}) and the classical
relation $\partial_j F^{(0)}_{ji}(\eta,\vec{x}) = -\epsilon^{ijk} m^j
\partial_k \delta^3(\vec{x})$ we infer,
\begin{equation}
J^{1K}_i(\eta,\vec{x}) = -\frac{\alpha H^4 a^2}{4 \pi^2}
\epsilon^{ijk} m^j \partial_k \int_{\eta_I}^{\eta-x} \!\!\!\!\!
d\eta' \, {a'}^2 \Biggl\{ \ln\Bigl[{H'}^2 (\Delta \eta^2 \!-\!
x^2)\Bigr] \!+\! 2\Biggr\} \; . \label{J1BK0}
\end{equation}

The next step is to perform the conformal time integrations in
expressions (\ref{J1BG0}) and (\ref{J1BK0}). For the $G$-type
current density comparison with (\ref{J1G01}) reveals that we
can read off the result by making the replacement $q \rightarrow
\epsilon^{ijk} m^j \partial_k$ in (\ref{J1G02}),
\begin{eqnarray}
\lefteqn{J^{1G}_{i}(\eta,\vec{x}) = \frac{\alpha H^2 a}{4 \pi^2}
\epsilon^{ijk} m^j \partial_k \Biggl\{ \frac{a_{-}}{x} \Biggl[
\ln\Bigl(\frac{a_{-}}{2 H x}\Bigr) \!-\! \ln\Bigl[1 \!-\!
\frac1{a_{-}}\Bigr] \!-\! \gamma \!-\! \frac12
\!+\! \frac1{2 a H x}\Biggr] } \nonumber \\
& & \hspace{2.5cm} + \frac{a_{+}}{x} \Biggl[ -\ln(a_{-}) \!-\! \ln(2
H x) \!+\! \ln\Bigl[1 \!-\! \frac1{a_{+}}\Bigr] \!+\! \frac12 \!-\!
\frac1{2 a H x}\Biggr] \Biggr\} . \qquad \label{J1BG1}
\end{eqnarray}
For the $K$-type current density it is best to change variables
from $\eta'$ to $a' = -1/H\eta'$ in expression (\ref{J1BK0}),
\begin{eqnarray}
\lefteqn{ J^{1K}_i(\eta,\vec{x}) = -\frac{\alpha H^3 a^2}{4 \pi^2}
\epsilon^{ijk} m^j \partial_k \!\! \int_{1}^{a_-} \!\!\!\!\! da' \,
\Biggl\{ \ln\Bigl[\frac1{a'} \!-\! \frac1{a_-} \Bigr] \!+\! \ln\Bigl[
\frac1{a'} \!-\! \frac1{a_+} \Bigr] \!+\! 2\gamma \Biggr\} \; , \qquad } \\
& & \hspace{-.3cm} = \frac{\alpha H^3 a^2}{4 \pi^2} \epsilon^{ijk}
m^j \partial_k \Biggl\{ (a_+ \!-\! a_-) \ln(a_+ \!-\! a_-) \!-\!
(a_- \!-\! 1) \ln(a_- \!-\! 1) \nonumber \\
& & \hspace{.8cm} - (a_+ \!-\! 1) \ln(a_+ \!-\! 1) \!-\! (1 \!-\!
a_-) \ln(a_+) \!+\! (1 \!+\! a_-) \ln(a_-) \!-\! 2 \gamma a_-
\Biggr\} \; . \qquad \label{J1BK1}
\end{eqnarray}

The final step before combining the $G$-type and $K$-type currents
is to put them in a common form by isolating distinct logarithms and
making judicious use of the identities,
\begin{equation}
2 a_{+} a_{-} = a (a_{+} \!+\! a_{-}) = \frac{(a_{+} \!-\! a_{-})}{H x}
\qquad , \qquad a H x a_{\pm} = \pm a_{\pm} \mp a \; . \label{goddIDs}
\end{equation}
The results are,
\begin{eqnarray}
\lefteqn{J^{1G}_{i}(\eta,\vec{x}) = \frac{\alpha H^2}{4 \pi^2}
\epsilon^{ijk} m^j \partial_k \Biggl\{ \frac{a}{x} \Biggl[
-( a_{+} \!+\! a_{-}) \ln(2 H x) \!-\! (a_{+} \!-\! a_{-})
\ln(a_{-}) } \nonumber \\
& & \hspace{3.5cm} - a_{-} \ln\Bigl(1 \!-\! \frac1{a_{-}}\Bigr) \!+\!
a_{+} \ln\Bigl(1 \!-\! \frac1{a_{+}} \Bigr) \!-\! (\gamma \!+\! 1) a_{-}
\Biggr] \Biggr\} \; , \qquad \label{J1BG2} \\
\lefteqn{J^{1K}_{i}(\eta,\vec{x}) = \frac{\alpha H^2}{4 \pi^2}
\epsilon^{ijk} m^j \partial_k \Biggl\{ \frac{a}{x} \Biggl[
(a_{+} \!+\! a_{-} \!-\! 2 a) \ln(2 H x) } \nonumber \\
& & \hspace{1cm} + (a_{+} \!+\! a_{-} \!+\!  2 a H x \!-\! 2 a)
\ln(a_{-}) \!+\! (a_{-} \!+\! a H x \!-\! a) \ln\Bigl( 1 \!-\!
\frac1{a_{-}}\Bigr) \nonumber \\
& & \hspace{4cm} - (a_{+} \!-\! a H x \!-\! a) \ln\Bigl(1 \!-\!
\frac1{a_{+}} \Bigr) \!+\! 2 \gamma (a_{-} \!-\! a) \Biggr] \Biggr\} \; .
\qquad \label{J1BK2}
\end{eqnarray}
Adding expressions (\ref{J1BG2}) and (\ref{J1BK2}) gives a current
density which is much simpler than either of them separately,
\begin{eqnarray}
\lefteqn{J^{GK}_{i}(\eta,\vec{x}) = \frac{\alpha H^2}{4 \pi^2}
\epsilon^{ijk} m^j \partial_k \Biggl\{ \frac{a}{x} \Biggl[
-2 a \ln(2 H x) \!+\! 2 (a_{-} \!+\! a H x \!-\! a) \ln(a_{-}) } \nonumber \\
& & \hspace{0cm} - a (1 \!-\! Hx) \ln\Bigl(1 \!-\! \frac1{a_{-}}\Bigr) \!+\!
a (1 \!+\! H x) \ln\Bigl(1 \!-\! \frac1{a_{+}} \Bigr) \!-\! a_{-}
\!+\! \gamma (a_{-} \!-\! 2 a) \Biggr] \Biggr\} \; . \qquad \label{J1BGK}
\end{eqnarray}
Finally, it is useful to introduce the symbol $J^{GK}(\eta,x)$ for
$H^2$ times the part of (\ref{J1BGK}) within the curly
brackets,
\begin{equation}
J^{GK}_{i}(\eta,\vec{x}) \equiv \frac{\alpha}{4 \pi^2}
\epsilon^{ijk} m^j \partial_k J^{GK}(\eta,x) \; . \label{JGK}
\end{equation}

From (\ref{dSeqn}) we see that the one loop field strength induced
by (\ref{J1BGK}) obeys the equation,
\begin{equation}
\partial^2 F^{GK}_{\mu\nu}(\eta,\vec{x}) = \partial_{\mu}
J^{GK}_{\nu}(\eta,\vec{x}) \!-\! \partial_{\nu} J^{GK}_{\mu}(\eta,\vec{x})
\; .
\end{equation}
The retarded solution (\ref{retsol}) takes the form,
\begin{equation}
F^{GK}_{\mu\nu}(\eta,\vec{x}) = \partial_{\mu}
\mathcal{J}^{GK}_{\nu}(\eta,\vec{x}) \!-\! \partial_{\nu}
\mathcal{J}^{GK}_{\mu}(\eta,\vec{x}) \; ,
\end{equation}
where,
\begin{equation}
\mathcal{J}^{GK}_{\mu}(\eta,\vec{x}) = -\int \!\! d^4x' \,
\frac{ \delta(\Delta \eta \!-\! \Delta x)}{4 \pi \Delta x} \,
J^{GK}_{\mu}(\eta',\vec{x}') \; .
\end{equation}
We can commute the differential operator in (\ref{JGK})
through the Greens function to define a scalar function
$\mathcal{J}^{GK}(\eta,x)$,
\begin{eqnarray}
\mathcal{J}^{GK}_i(\eta,\vec{x}) & = & \frac{\alpha}{4 \pi^2}
\epsilon^{ijk} m^j \partial_k \mathcal{J}^{GK}(\eta,x) \; , \\
& \equiv & -\frac{\alpha}{4 \pi^2} \epsilon^{ijk} m^j \partial_k
\int \!\! d^4x' \, \frac{ \delta(\Delta \eta \!-\! \Delta x)}{4
\pi \Delta x} \, J^{GK}(\eta',x') \; . \qquad \label{scriptJGK}
\end{eqnarray}
In terms of $\mathcal{J}^{GK}(\eta,x)$ the electric and magnetic
field strengths are,
\begin{eqnarray}
F^{GK}_{0i}(\eta,\vec{x}) & = & \frac{\alpha \epsilon^{ijk}}{4 \pi^2}
\, m^j \partial_k \partial_0 \mathcal{J}^{GK}(\eta,x) \; , \\
F^{GK}_{ij}(\eta,\vec{x}) & = & \frac{\alpha \epsilon^{ijk}}{4 \pi^2}
\Bigl[m^k \nabla^2 \!-\! \vec{m} \!\cdot\! \vec{\nabla} \partial^k
\Bigr] \mathcal{J}^{GK}(\eta,x) \; .
\end{eqnarray}

It remains to evaluate the function $\mathcal{J}^{GK}(\eta,x)$.
Note that for a function $f(\eta,x)$ of just $\eta$ and $x \equiv
\Vert \vec{x}\Vert$, the integral against the retarded Green's
function gives,
\begin{equation}
-\int \!\! d^4x' \, \frac{\delta(\Delta \eta \!-\! \Delta
x)}{4\pi \Delta x} \times f(\eta',x') = -\frac1{2 x}
\int_{\eta_I}^{\eta} \!\!\! d\eta' \! \int_{\vert x - \Delta
\eta\vert}^{x + \Delta \eta} \!\!\!\!\!\!\!\!\!\!\! dx' x'
f(\eta',x') \; .
\end{equation}
In our case the function $f(\eta,x)$ is $J^{GK}(\eta,x)$ and
we must remember that causality makes it vanish for $x > \eta - \eta_I$.
We can therefore express $\mathcal{J}^{GK}(\eta,x)$ as the
sum of three integrals,
\begin{eqnarray}
\lefteqn{\mathcal{J}^{GK}(\eta,x) = -\frac1{2x} \Biggl\{
\int_{\frac{\eta + \eta_I -x}2}^{\eta - x} \!\!\!\! d\eta' \!\!
\int_{\eta - \eta' - x}^{\eta' - \eta_I} \!\!\!\! dx'
\!+\! \int_{\eta - x}^{\frac{\eta + \eta_I + x}2} \!\!\!\! d\eta' \!\!
\int_{x - \eta + \eta'}^{\eta' - \eta_I} \!\!\!\! dx' } \nonumber \\
& & \hspace{5.5cm} + \int_{\frac{\eta + \eta_I +x}2}^{\eta} \!\!\!\!
d\eta' \!\! \int_{x - \eta + \eta'}^{x + \eta - \eta'} \!\!\!\! dx'
\Biggr\} \, x' J^{GK}(\eta',x') \; . \qquad \label{longint}
\end{eqnarray}
These integrals are evaluated in Appendix B. Here we present just
the leading results for large $a$ with $x$ fixed, and with $r = x/a$
fixed,
\begin{equation}
\mathcal{J}^{GK}_{{\rm fixed} \, x} \longrightarrow \frac{\ln(a)}{x}
\times \mathcal{G}(H x) \qquad , \qquad \mathcal{J}^{GK}_{{\rm
fixed} \, r} \longrightarrow \frac{\ln(a)}{x} \times \mathcal{K}(H r) 
\; ,
\end{equation}
where the functions $\mathcal{G}(z)$ and $\mathcal{K}(z)$ are,
\begin{eqnarray}
\mathcal{G}(z) & \equiv & 2 \gamma \!+\! 2 \ln(2) \!-\!
\ln\Bigl(\frac{1 \!+\! z}{1 \!-\! z}\Bigr) \!-\!
z \ln(1 \!-\! z^2) \; , \qquad \label{Gdef} \\
\mathcal{K}(z) & \equiv & \frac{\pi^2}{12} \!+\! 2 \ln(1 \!+\! z)
\!+\! {\rm Li}_2\Bigl( \frac{z \!-\! 1}{z \!+\! 1}\Bigr) \; .
\label{Kdef}
\end{eqnarray}

The function $\mathcal{G}(z)$ is positive at $z = 0$ so the $GK$
terms dominate over the $F$ and $C$ terms at late times for fixed
$x$,
\begin{eqnarray}
\Bigl( F^{(1)}_{0i} \Bigr)_{{\rm fixed} \, x} & \longrightarrow &
\frac{\alpha \epsilon^{ijk}}{4\pi^2} \, m^j \partial_k \partial_0
\Biggl\{ \frac{\ln(a)}{x} \times \mathcal{G}(H x) \Biggr\} \; , \\
\Bigl( F^{(1)}_{ij} \Bigr)_{{\rm fixed} \, x} & \longrightarrow &
\frac{\alpha \epsilon^{ijk}}{4\pi^2} \, \Bigl[m^k \nabla^2 \!-\!
\vec{m} \!\cdot\! \vec{\nabla} \partial_k \Bigr] \Biggl\{
\frac{\ln(a)}{x} \times \mathcal{G}(H x) \Biggr\} \; .
\label{B1late}
\end{eqnarray}
Near $x=0$ the term in curly brackets is the positive constant $2
\gamma + 2 \ln(2)$ times $\ln(a)/x$. This screens the classical
result of $-1/x$, although only logarithmically.

For small $z$ the function $\mathcal{K}(z)$ has the expansion,
\begin{equation}
\mathcal{K}(z) = 2 \ln(2) z - 2 z^2 + O(z^3) \; .
\end{equation}
The linear term drops out from the one loop field strengths, so the
$GK$ terms are negligible compared with the $F$ and $C$ terms for
large $a$ at fixed $r$. The resulting field strengths are,
\begin{eqnarray}
\lefteqn{\Bigl( F^{(1)}_{0i} \Bigr)_{{\rm fixed} \, r} \longrightarrow
\frac{\alpha \epsilon^{ijk}}{24\pi^2} \, m^j \partial_k \partial_0
\Biggl\{ \frac{-\ln(2\mu x) \!+\! \ln(Hx \!+\! \frac1{a})}{x}
\Biggr\} \; , \qquad } \\
& & \hspace{4cm} = \frac{\alpha \epsilon^{ijk}}{24 \pi^2} \, m^j
\widehat{r}^k \times H a^3 \times \Biggl\{ \frac1{r^2} \!-\!
\frac{H^2}{(1 \!+\! H r)^2} \Biggr\} \; , \qquad \label{E1atr} \\
\lefteqn{\Bigl( F^{(1)}_{ij} \Bigr)_{{\rm fixed} \, r} \longrightarrow
\frac{\alpha \epsilon^{ijk}}{24\pi^2} \, \Bigl[m^k \nabla^2 \!-\!
\vec{m} \!\cdot\! \vec{\nabla} \partial_k \Bigr] \Biggl\{
\frac{-\ln(2\mu x) \!+\! \ln(Hx \!+\! \frac1{a})}{x} \Biggr\} \; ,} \\
& & = \frac{\alpha \epsilon^{ijk}}{24 \pi^2} \times a^3
\times \Biggl\{ \ln\Bigl( \frac{1 \!+\! Hr}{2 \mu r}\Bigr) \Bigl[
\frac{m^k \!-\! 3 \vec{m} \!\cdot\! \widehat{r} \widehat{r}^k}{r^3}
\Bigr] \!+\! \frac{2}{r^3} \Bigl[ m^k \!-\! 2 \vec{m} \!\cdot\!
\widehat{r} \widehat{r}^k \Bigr] \nonumber \\
& & \hspace{4cm} - \frac{H}{r^2} \Bigl[ \frac{m^k \!-\! 3 \widehat{m}
\!\cdot\! \widehat{r} \widehat{r}^k}{1 \!+\! H r} \Bigr] \!-\!
\frac{H^2}{r} \Bigl[ \frac{m^k \!-\! \widehat{m} \!\cdot\! \widehat{r}
\widehat{r}^k}{ (1 \!+\! H r)^2} \Bigr] \Biggr\} \; . \qquad \label{B1atr}
\end{eqnarray}

We must employ expressions (\ref{staticE}) and (\ref{staticB}) to
convert (\ref{E1atr}) and (\ref{B1atr}) into the static field strengths.
The final result for the (late time form of the) static electric field
is relatively simple,
\begin{equation}
\widetilde{F}^{(1)}_{0i} \longrightarrow \frac{\alpha
\epsilon^{ijk}}{24\pi^2} \, m^j \widehat{r}^k \times H e^{Ht}
\sqrt{1 \!-\! H^2 r^2} \, \Biggl\{ -\frac{\ln(\frac{1 + Hr}{2\mu r}) + 1}{r^2}
+ \frac{H}{r (1 \!+\! Hr)} \Biggr\} \; . \label{EM1tilde}
\end{equation}
The factor of $e^{Ht}$ makes this seem like a substantial modification
but one should keep in mind that the tree order result carries the
same factor in static coordinates,
\begin{equation}
\widetilde{F}^{(0)}_{0i} = \frac{\epsilon^{ijk}}{4\pi} \, m^j \widehat{r}^k
\times -\frac{H e^{Ht} \sqrt{1 \!-\! H^2 r^2}}{r^2} \; . \label{EM0tilde}
\end{equation}
It is therefore better to view the one loop correction as a time
independent screening of the classical result by the $r$-dependent
enhancement (for small $r$) of about $\alpha/6\pi \times \ln(1/2 \mu r)$.
This is exactly what one finds in flat space (\ref{1loopB}) and is a
manifestation of the running of the electrodynamic coupling. The tensor
structure of the one loop static magnetic field is complicated but
its leading term for small $r$ bears the same relation to the tree
order result,
\begin{eqnarray}
\widetilde{F}^{(0)}_{ij} & \!\!\!=\!\!\! & \frac{\epsilon^{ijk}}{4 \pi}
\times e^{Ht} \sqrt{1 \!-\! H^2 r^2} \, \Biggl\{ \Bigl[ \frac{ m^k \!-\!
3 \vec{m} \!\cdot\! \widetilde{r} \widetilde{r}^k}{r^3}\Bigr] \!-\!
\frac{H^2}{r} \Bigl[ \frac{ m^k \!-\! \vec{m} \!\cdot\! \widetilde{r}
\widetilde{r}^k}{1 \!-\! H^2 r^2}\Bigr] \Biggr\} , \qquad \label{B0tilde} \\
\widetilde{F}^{(1)}_{ij} & \!\!\!\longrightarrow \!\!\!& \frac{\alpha
\epsilon^{ijk}}{24 \pi^2} \times e^{Ht} \sqrt{1 \!-\! H^2 r^2} \,
\Biggl\{ \ln\Bigl(\frac{1 \!+\! Hr}{2\mu r} \Bigr) \Bigl[ \frac{ m^k \!-\!
3 \vec{m} \!\cdot\! \widetilde{r} \widetilde{r}^k}{r^3}\Bigr] + \dots
\Biggr\} . \label{B1tilde}
\end{eqnarray}

\section{Discussion}

In this study we have explored the effect that the inflationary
production of charged, massless and minimally coupled scalars have
on the electric and magnetic fields induced by a point charge
and by a point magnetic dipole. The sources were held fixed at
the spatial origin. We derived results for the fields perceived
by two observers, one stationary in conformal coordinates and the
other stationary in static coordinates.

The conformal observer sees the one loop correction (\ref{E1late})
to the electric field of a point charge screen the classical result
exponentially rapidly in physical time. That is to be expected
because the physical distance between this observer and the source
grows exponentially, and the intervening space is filled with a
nearly constant density of charged particles. In contrast, the
static observer perceives the one loop correction (\ref{E1tilde})
to slowly {\it enhance} the classical result. It is more difficult
to understand this result. A possible interpretation is that it
derives from the running of the electrodynamic coupling which is
built into the scalar vacuum. However, it should be noted that
the enhancement does not come from merely probing the one loop
correction in flat space (\ref{Efield}) at the ever-smaller
coordinate separations needed to keep the physical distance to
the source constant. That effect is entirely canceled by the
conformal anomaly term (\ref{E1C}); the enhancement originates
in the intrinsically de Sitter contribution (\ref{fullE1G}).

Our results for a current dipole are interestingly different.
The conformal observer sees much weaker screening (\ref{B1late})
which grows only linearly in physical time. The static observer
sees an enhancement (\ref{B1tilde}) which is constant in time but
grows as one approaches the source. Because the static result
derives from the flat space and conformal anomaly terms it is the
most simply understood in terms of the flat space result
(\ref{1loopB}), and is clearly a manifestation of the running of
the electrodynamic coupling.

As far as we know this work is unique in having solved the
effective field equations for the constrained part of a force
field on de Sitter background. There have been many previous
studies of the propagation of dynamical quanta on de Sitter,
both from the production of scalars \cite{phi4,PTW,PW2,Yukawa,QG}
and from the production of gravitons \cite{Dirac1,Dirac2,MMCS}. There
have also been many flat space background studies of how massless
scalars affect the force of gravity \cite{many,SPW}, and of how
gravitons affect electric and magnetic forces \cite{few,LW}. But
no one has previously combined de Sitter and force laws to reveal
the surprising dichotomy between screening in conformal coordinates
to anti-screening in static coordinates.

Although this seems to be the first time one loop corrections to
force laws have been studied on de Sitter, it will not be the last.
Computations have already been made of the one loop graviton
self-energy from inflationary gravitons \cite{TW2} and from
inflationary scalar \cite{SPW2}, and the vacuum polarization from
inflationary gravitons will soon be completed \cite{LW2}. These
results can be used to quantum-correct the effective Maxwell and
Einstein equations, and then proceed to a study of the same type we
have just completed. We anticipate that the current work will play
an important role in organizing and understanding these studies. The
project to quantum-correct Maxwell's equations to account for
inflationary gravitons is especially noteworthy because it is a
straightforward approach to check the puzzling screening effect that
Kitamoto and Kitazawa have inferred from graviton loop corrections
to the off-shell effective field equations at fixed sub-horizon
scales \cite{KK}.

\vskip .5cm

\centerline{\bf Acknowledgements}

This work was partially supported by the master student exchange
program at the \'Ecole Polytechnique F\'ed\'erale de Lausanne, by
NSF grants PHY-0855021, PHY-1205591, and by the Institute for
Fundamental Theory at the University of Florida. We are grateful to
M. Shaposhnikov for many valuable comments.

\section{Appendix A: Integrals from Subsection \ref{pointQ}}

The following integrals (times $-\alpha/\pi \times F^{(0)}_{0i}$)
sum to give our result (\ref{fullE1G}) for $F^{1G}_{0i}$ in section
4.2,
\begin{eqnarray}
\lefteqn{a H^2 \! \int_0^x \!\! dx' x' a_{-}' \ln(a_{-}') = \frac12
\ln^2(a_{-}) \!-\! \frac12 \ln^2(a) \!+\! \frac{a}{a_{-}} \ln(a_{-})
\!-\! \ln(a) \!+\! a H x \; ,} \label{Int1} \\
\lefteqn{-a H^2 \! \int_0^x \!\! dx' x' a_{-}' \ln(2 H x') = \ln(1
\!+\! a H x) \ln(2 H x) \!+\! {\rm Li}_2(-a H
x) \; ,} \nonumber \\
& & \hspace{7.5cm} -a H x \ln(2Hx) \!+\! a H x \; , \qquad
\label{Int2} \\
\lefteqn{-a H^2 \! \int_0^x \!\! dx' x' a_{-}' \ln\Bigl(1 \!-\!
\frac1{a_{-}'}\Bigr) = \Bigl(a \!-\! \frac{a}{a_{-}}\Bigr)
\ln\Bigl(1 \!-\! \frac1{a_{-}}\Bigr) \!-\! \Bigl(a \!-\! 1\Bigr)
\ln\Bigl(1 \!-\! \frac1{a}\Bigr) } \nonumber \\
& & \hspace{7cm} + {\rm Li}_2\Bigl(\frac1{a}\Bigr) \!-\! {\rm
Li}_2\Bigl(\frac1{a_{-}}\Bigr) \!+\! a H x \; , \qquad \label {Int3} \\
\lefteqn{-\Bigl(\gamma \!+\! \frac12\Bigr) a H^2 \! \int_0^x \!\!
dx' x' a_{-}' = \Bigl(\gamma \!+\! \frac12\Bigr) \Bigl[ \ln(a) \!-\!
\frac{a}{a_{-}} \ln(a_{-}) \!-\! a H x\Bigr] \; ,} \label{Int4} \\
\lefteqn{a H^2 \! \int_0^x \!\! dx' \frac{a_{-}'}{2 a H} =
\frac12 \ln(a) \!-\! \frac12 \ln(a_{-}) \; , } \label{Int5} \\
\lefteqn{-a H^2 \! \int_0^x \!\! dx' x' a_{+}' \ln(a_{-}') =
\ln\Bigl(\frac{a}2\Bigr) \ln(1 \!-\! a H x) \!+\! \frac{a}{a_{-}}
\ln(a_{-}) \!-\! \ln(a) \!+\! a H x \; ,} \nonumber \\
& & \hspace{6.5cm} - {\rm Li}_2\Bigl(\frac12\Bigr) \!+\! {\rm
Li}_2\Bigl(\frac12 \!-\! \frac12 a H x\Bigr) \; , \label{Int6} \\
\lefteqn{-a H^2 \! \int_0^x \!\! dx' x' a_{+}' \ln(2 H x') =
-\ln\Bigl( \frac{a}2 \Bigr) \ln(1 \!-\! a H x) \!+\! a H x \ln(2 H
x) \!-\! a H x} \nonumber \\
& & \hspace{7cm} + {\rm Li}_2(1) \!-\! {\rm Li}_2(1 \!-\! a H x)
\; , \label{Int7} \\
\lefteqn{a H^2 \! \int_0^x \!\! dx' x' a_{+}' \ln\Bigl(1 \!-\!
\frac1{a_{+}'}\Bigr) = -\Bigl(a \!-\! \frac{a}{a_{+}}\Bigr)
\ln\Bigl(1 \!-\! \frac1{a_{+}}\Bigr) \!+\! \Bigl(a \!-\! 1\Bigr)
\ln\Bigl(1 \!-\! \frac1{a}\Bigr) } \nonumber \\
& & \hspace{7cm} - {\rm Li}_2\Bigl(\frac1{a}\Bigr) \!+\!
{\rm Li}_2\Bigl(\frac1{a_{+}}\Bigr) \!+\! a H x \; , \qquad
\label{Int8} \\
\lefteqn{\frac12 a H^2 \! \int_0^x \!\! dx' x' a_{+}' = -\frac12
\ln(1 \!-\! a H x) \!-\! \frac12 a H x \; , } \label{Int9} \\
\lefteqn{-a H^2 \! \int_0^x \!\! dx' \frac{a_{+}'}{2 a H} = \frac12
\ln(1 \!-\! a H x) \; .} \label{Int10}
\end{eqnarray}
The integrands of (\ref{Int6}-\ref{Int7}) and
(\ref{Int9}-\ref{Int10}) diverge at $x' = -\eta$, which results in
the ill-defined factors of $\pm \ln(1 - a H x)$. However, these
divergences cancel between (\ref{Int6}-\ref{Int7}) and between
(\ref{Int9}-\ref{Int10}), as do the problematic logarithms.

\section{Appendix B: Integrals from Subsection \ref{pointM}}

We begin with some general comments. The expression (\ref{longint})
we must evaluate is a double integration over three regions which
can be termed ``$I$'', ``$II$'' and ``$III$'',
\begin{equation}
\mathcal{J}^{GK}(\eta,x) = -\frac1{2x} \Biggl[ \int_{I} \!\! d\eta'
dx' \!+\! \int_{II} \!\!\! d\eta' dx' \!+\! \int_{III} \!\!\!\!
d\eta' dx' \Biggr] x' J^{GK}(\eta',x') \; .
\end{equation}
Each of these double integrals takes the form,
\begin{equation}
\int_{B}^{A} \!\! d\eta' \! \int_{D(\eta')}^{C(\eta')} \!\! dx' \; ,
\end{equation}
where the various limits are,
\begin{eqnarray}
A_I = \eta \!-\! x & , & A_{II} = \frac12 \Bigl( \eta_I \!+\! \eta
\!+\! x\Bigr) \quad , \quad A_{III} = \eta \; , \\
B_I = \frac12 \Bigl( \eta_I \!+\! \eta \!-\! x\Bigr) & , & B_{II} =
A_I \quad , \quad B_{III} = A_{II} \; , \\
C_I = \eta' \!-\! \eta_I & , & C_{II} = C_I \quad , \quad C_{III} =
\eta \!+\! x \!-\! \eta' \; , \\
D_I = \eta \!-\! x \!-\! \eta' & , & D_{II} = -D_I \quad , \quad
D_{III} = -D_I \; .
\end{eqnarray}

Of course the integrand is of crucial importance. From expressions
(\ref{J1BGK}) and (\ref{JGK}) we see that $x J^{GK}(\eta,x)$ can be
broken up into seven components,
\begin{eqnarray}
\mathcal{I}_1(\eta,x) & = & -2 H^2 a^2 \ln(2 H x) \; , \label{I1} \\
\mathcal{I}_2(\eta,x) & = & +2 H^2 a a_{-} \ln(a_{-}) \; , \label{I2} \\
\mathcal{I}_{34}(\eta,x) & = & +2 H^2 a^2 (1 \!-\! H x) \ln\Bigl[-H
(\eta \!-\! x)\Bigr] \; , \label{I34} \\
\mathcal{I}_{56}(\eta,x) & = & -a^2 (1 \!-\! H x) \ln\Bigl[1 \!+\! H
(\eta \!-\! x)\Bigr] \; , \label{I56} \\
\mathcal{I}_{78}(\eta,x) & = & +a^2 (1 \!+\! H x) \ln\Bigl[1 \!+\! H
(\eta \!+\! x)\Bigr] \; , \label{I78} \\
\mathcal{I}_9(\eta,x) & = & (\gamma \!-\! 1) a a_{-} \; , \label{I9} \\
\mathcal{I}_{10}(\eta,x) & = & -2 a^2 \; . \label{I10}
\end{eqnarray}
In each case the $x'$ integration can be expressed in terms of
elementary functions. For $\mathcal{I}_{10}$ the $\eta'$ integration
also results in elementary functions. $\mathcal{I}_2$ and $\mathcal{I}_9$
give polylogarithms, about which more later. The remaining integrands
--- $\mathcal{I}_1$, $\mathcal{I}_{34}$, $\mathcal{I}_{56}$ and
$\mathcal{I}_{78}$ --- all take the form of ${a'}^2$ times the $x'$
derivative of an elementary function of $\eta'$ and $x'$. For these
integrands the best strategy is partially integrate on $\eta'$ after
having performed the $x'$ integration,
\begin{eqnarray}
\lefteqn{H \!\! \int_{B}^{A} \!\!\! d\eta' {a'}^2 \!\!\!
\int_{D(\eta')}^{C(\eta')} \!\!\!\!\!\!\!\! dx' \,
\frac{\partial F(\eta',x')}{\partial x'} = H \!\! \int_{B}^{A} \!\!\!
d\eta' {a'}^2 \Biggl\{ F\Bigl(\eta', C(\eta')\Bigr) \!-\!
F\Bigl(\eta',D(\eta')\Bigr) \Biggr\} \; , \qquad } \\
& & = a' \Biggl\{ F\Bigl(\eta' , C\Bigr) \!-\!
F\Bigl(\eta' , D\Bigr) \Biggr\} \Biggr\vert_{B}^{A} \!-\!
\int_{B}^{A} \!\!\! d\eta' a' \frac{\partial}{\partial \eta'} \Biggl\{
F\Bigl(\eta' , C\Bigr) \!-\! F\Bigl(\eta' , D\Bigr)
\Biggr\} \; . \qquad \label{genstrat}
\end{eqnarray}
It turns out that the surface terms in (\ref{genstrat}) cancel
between the three regions,
\begin{eqnarray}
\lefteqn{a' \Biggl\{ F\Bigl(\eta' , C_{I}\Bigr) \!-\!
F\Bigl(\eta' , D_{I}\Bigr) \Biggr\} \Biggr\vert_{B_I}^{A_I} \!\!\!\! =
a_{A_I} \Biggl\{ F\Bigl( A_I, \eta \!-\! x \!-\! \eta_I\Bigr)
\!-\! F\Bigl(A_I,0\Bigr) \Biggr\} \!-\! 0 \; , \qquad } \\
\lefteqn{ a' \Biggl\{ F\Bigl(\eta' , C_{II}\Bigr) \!-\!
F\Bigl(\eta' , D_{II}\Bigr) \Biggr\} \Biggr\vert_{B_{II}}^{A_{II}}
\!\!\!\!= a_{A_{II}} \Biggl\{ F\Bigl( A_{II}, \frac12 (\eta \!+\! x
\!-\! \eta_I)\Bigr) } \nonumber \\
& & \hspace{.5cm} - F\Bigl( A_{II}, \frac12 (\eta_I \!-\! \eta \!+\!
3x )\Bigr) \Biggr\} \!-\! a_{A_I} \Biggl\{ F\Bigl( A_I, \eta \!-\! x
\!-\! \eta_I\Bigr) \!-\! F\Bigl(A_I,0\Bigr) \Biggr\} \; , \qquad \\
\lefteqn{ a' \Biggl\{ F\Bigl(\eta' , C_{III}\Bigr) \!-\!
F\Bigl(\eta' , D_{III}\Bigr) \Biggr\} \Biggr\vert_{B_{III}}^{A_{III}}
\!\!\!\! = 0 \!-\! a_{A_{II}} \Biggl\{ F\Bigl( A_{II}, \frac12 (\eta
\!+\! x \!-\! \eta_I )\Bigr) } \\
& & \hspace{7cm} - F\Bigl( A_{II}, \frac12 (\eta_I \!-\! \eta \!+\! 3x
)\Bigr) \Biggr\} \; . \qquad
\end{eqnarray}

Integrands $\mathcal{I}_2$ and $\mathcal{I}_9$, and the volume terms
(\ref{genstrat}) from $\mathcal{I}_1$, $\mathcal{I}_{34}$, $\mathcal{I}_{56}$
and $\mathcal{I}_{78}$, all give rise to $\eta'$ integrations of the
form $1/\eta'$ times logarithms. Integrations of this form result
in polylogarithms, of which the two we require are,
\begin{equation}
{\rm Li}_2(z) \equiv -\int_0^z \!\! dt \, \frac{\ln(1 \!-\! t)}{t}
\qquad , \qquad {\rm Li}_3(z) \equiv \int_0^z \!\! dt \,
\frac{ {\rm Li}_2(t)}{t} \; .
\end{equation}
Both are real for $z \leq 1$. Their expansions for small $z$ and for
large $-z$ are,
\begin{eqnarray}
{\rm Li}_2(z) \longrightarrow z + O(z^2) & , & {\rm Li}_2(z)
\longrightarrow -\frac12 \ln^2(-z) -\frac{\pi^2}{6} +
O\Bigl(\frac1{z} \Bigr) \; , \qquad \\
{\rm Li}_3(z) \longrightarrow z + O(z^2) & , & {\rm Li}_3(z)
\longrightarrow -\frac16 \ln^3(-z) - \frac{\pi^2}{6} \ln(-z) +
O\Bigl(\frac1{z} \Bigr) \; . \qquad
\end{eqnarray}
Care must be taken to arrange things so that the arguments
of ${\rm Li}_2(z)$ and ${\rm Li}_3(z)$ lie in the range $z \leq 1$
for which the function is real. This is simple to accomplish for the
dilogarithm ${\rm Li}_2(z)$ but it requires some effort for
${\rm Li}_3(z)$. ${\rm Li}_3(z)$ derives exclusively from $\mathcal{I}_2$,
when changing to dimensionless variables converts the $\eta'$
integration to either $\int dt/t \times \ln^2(1 - t)$ or $\int dt/t
\times \ln(t) \ln(1 - t)$. The final integral only makes sense for
$0 \leq t \leq 1$ so it involves no choices,
\begin{equation}
\int \!\! dt \, \frac{\ln(t) \ln(1 \!-\! t)}{t} = {\rm Li}_3(t)
\!-\! \ln(t) {\rm Li}_2(t) \; .
\end{equation}
If the dimensionless parameter $t$ lies in the range $0 \leq t \leq 1$
we write the first integral as,
\begin{equation}
\int \!\! dt \frac{\ln^2(1 \!-\! t)}{t} = -2 {\rm Li}_3(1 \!-\! t)
\!+\! 2 \ln(1 \!-\! t) {\rm Li}_2(1 \!-\! t) \!+\! \ln^2(1 \!-\! t)
\ln(t) \; . \label{1stform}
\end{equation}
However, if $t < 0$ we must use Landen's identity to re-express it as,
\begin{eqnarray}
\lefteqn{\int \!\! dt \frac{\ln^2(1 \!-\! t)}{t} = -2 {\rm Li}_3(t)
\!-\! 2 {\rm Li}_3\Bigl( \frac{-t}{1 \!-\! t}\Bigr) \!-\! 4
{\rm Li}_3\Bigl(\frac1{1 \!-\! t}\Bigr) \!+ 2 \ln\Bigl(
\frac1{1 \!-\! t}\Bigr) {\rm Li}_2\Bigl(\frac{-t}{1 \!-\! t}\Bigr) }
\nonumber \\
& & \hspace{3.8cm} + 4 \ln\Bigl(\frac1{1 \!-\! t}\Bigr)
{\rm Li}_2\Bigl(\frac1{1 \!-\! t}\Bigr) \!+\! 2 \ln^2\Bigl(\frac1{1
\!-\! t}\Bigr) \ln\Bigl(\frac{-t}{1 \!-\! t} \Bigr) \; . \qquad
\label{2ndform}
\end{eqnarray}
It can happen that the appropriate choice to make between (\ref{1stform})
and (\ref{2ndform}) depends upon whether $x$ or $r = a x$ is held fixed
for large $a$, in which case we report the choice appropriate for
holding $x$ fixed at large $a$.

Our results for the seven integrals are,
\begin{eqnarray}
\lefteqn{-2 x \mathcal{J}^{GK}_1 = 2 {\rm Li}_2\Bigl(1 \!-\!
a H x\Bigr) \!-\! 2 {\rm Li}_2\Bigl( \frac1{1 \!+\! a Hx}\Bigr) \!+\!
2 {\rm Li}_2\Bigl[ \frac12 \Bigl(1 \!+\! \frac1{a} \!-\! Hx\Bigr)
\Bigr] } \nonumber \\
& & \hspace{-.5cm} -2 {\rm Li}_2\Bigl[\frac12 \Bigl(1 \!+\! \frac1{a} \!+\!
Hx\Bigr) \Bigr] \!+\! 2 {\rm Li}_2\Bigl[2 \Bigl( \frac{1 \!+\! a H x}{a \!+\!
1 \!+\! a H x}\Bigr)\Bigr] \!-\! 2 {\rm Li}_2\Bigl[2 \Bigl(
\frac{1 \!-\! a H x}{a \!+\! 1 \!-\! a H x}\Bigr) \Bigr] \qquad \nonumber \\
& & \hspace{-.5cm} + \ln^2\Bigl( \frac{a}2\Bigr) \!-\! \ln\Bigl(1 \!+\!
a H x\Bigr) \ln\Bigl[2 \Bigl( \frac1{a} \!+\! Hx\Bigr) \Bigr] \!+\!
\ln^2\Bigl[2 \Bigl(1 \!+\! \frac1{a} \!+\! Hx\Bigr)\Bigr] \qquad \nonumber \\
& & \hspace{8cm} - \ln^2\Bigl[2 \Bigl(1 \!+\! \frac1{a} \!-\! Hx\Bigr)\Bigr]
\; , \qquad \\
\lefteqn{-2 x \mathcal{J}^{GK}_2 = 2 {\rm Li}_3\Bigl(
\frac{-a}{a H x \!+\! 1} \Bigr) \!-\! 2 {\rm Li}_3\Bigl( \frac{a}{a H x \!-\! 1}
\Bigr) \!+\! 2 {\rm Li}_3\Bigl(\frac{a H x}{a H x \!-\! 1}\Bigr) \!-\!
2 {\rm Li}_3(-1) } \nonumber \\
& & \hspace{-.5cm} + 2 \ln\Bigl( \frac1{a} \!+\! Hx\Bigr) \Biggl[
{\rm Li}_2(-1) \!-\! {\rm Li}_2\Bigl( \frac{a H x}{a H x \!-\! 1} \Bigr)
\Biggr] \!+\! \ln^2\Bigl( \frac1{a} \!+\! Hx\Bigr) \ln\Bigl(
\frac{1 \!-\! a H x}{1 \!+\! a H x}\Bigr) \; , \qquad \\
\lefteqn{-2 x \mathcal{J}^{GK}_{34} = \Bigl[4 \!+\! \frac2{a} \!+\! 2 H x\Bigl]
\Biggl\{ - {\rm Li}_2\Bigl( \frac{1 \!-\! a H x}{2} \Bigr) \!+\!
{\rm Li}_2\Bigl( \frac12 \Bigr) \!-\! {\rm Li}_2\Bigl(\frac{1 \!+\! a H x}{
a \!+\! 1 \!+\! a H x }\Bigr) } \nonumber \\
& & \hspace{-.5cm} + {\rm Li}_2\Bigl( \frac{1 \!-\! a H x}{a \!+\! 1 \!-\!
a H x}\Bigr) \!+\! \frac12 \ln^2\Bigl( \frac{1 \!+\! a H x}{2 a}\Bigr) \!-\!
\frac12 \ln^2\Bigl(\frac{a}2\Bigr) \!+\! \frac12 \ln^2\Bigl(1 \!+\! \frac1{a}
\!-\! Hx\Bigr) \qquad \nonumber \\
& & \hspace{-.5cm} - \frac12 \ln^2\Bigl(1 \!+\! \frac1{a} \!+\! Hx\Bigr)
\Biggr\} \!+\! \frac4{a} \ln\Bigl(a \!+\! 1 \!-\! a H x\Bigr) \!+\! 2
\Bigl(1 \!+\! \frac1{a} \!+\! Hx\Bigr) \ln\Bigl( \frac{1 \!+\!
a H x}{a \!+\! 1 \!+\! a H x}\Bigr) \nonumber \\
& & \hspace{-.5cm} - 2 \Bigl(1 \!-\! \frac1{a} \!-\! Hx\Bigr) \ln\Bigl(
\frac{1 \!+\! a H x}{a \!+\! 1 \!-\! a H x}\Bigr) \!-\! 2 \Bigl(\frac1{a}
\!+\! H x\Bigr) \ln\Bigl( \frac1{a} \!+\! Hx\Bigr) \ln\Bigl( 1 \!+\!
a H x\Bigr) \; , \qquad \\
\lefteqn{-2 x \mathcal{J}^{GK}_{56} = \Bigl[1 \!+\! \frac1{a} \!+\!
H x\Bigl] \Biggl\{ {\rm Li}_2\Bigl( \frac{2 \!+\! 2 a H x}{a \!+\! 1
\!+\! a H x}\Bigr) \!-\! {\rm Li}_2(1) \!+\! \ln\Bigl( \frac{a \!+\!
1 \!+\! a H x}{2 + 2 a H x}\Bigr) } \nonumber \\
& & \hspace{-.5cm} + \ln\Bigl( 1 \!+\! \frac1{a} \!+\! H x\Bigr)
\ln\Bigl( \frac{a \!+\! 1 \!+\! a H x}{2 \!+\! 2 a H x}\Bigr)
\Biggr\} \!-\! \Bigl[1 \!+\! \frac1{a} \!-\! Hx\Bigr] \Biggl\{ {\rm
Li}_2\Bigl(\frac{2}{a \!+\! 1 \!-\! a H x }\Bigr) \nonumber \\
& & \hspace{-.5cm} - {\rm Li}_2(1) \!+\! \ln\Bigl( \frac{a \!+\! 1
\!-\! a H x}{2 \!+\! 2 a H x}\Bigr) \!+\! \ln\Bigl(1 \!-\! \frac1{a}
\!-\! H x\Bigr) \ln\Bigl(1 \!+\! a H x\Bigr) \Biggr\} \nonumber \\
& & \hspace{-.5cm} -\frac{2}{a} \ln\Bigl(\frac{a \!+\! 1 \!-\! a H
x}{2}\Bigr) \!-\! \Bigl(1 \!+\! \frac1{a} \!-\! H x\Bigr) \ln\Bigl(1
\!+\! \frac1{a} \!-\! H x\Bigr) \ln\Bigl( \frac{a \!+\! 1 \!-\! a H
x}{2}\Bigr) \; , \qquad \\
 \lefteqn{-2 x \mathcal{J}^{GK}_{78} = \Bigl[1 \!+\! \frac1{a}
\!+\! H x\Bigl] \Biggl\{ {\rm Li}_2\Bigl( \frac{2}{a \!+\! 1 \!+\! a
H x}\Bigr) \!-\! {\rm Li}_2\Bigl( \frac{2 \!+\! 2 a H x}{a \!+\! 1
\!+\! a H x}\Bigr) } \nonumber \\
& & \hspace{-.5cm} - \ln\Bigl( \frac{a \!+\! 1 \!+\! a H x}{2 \!+\!
2 a H x}\Bigr) \!+\! \ln\Bigl(1 \!+\! \frac1{a} \!+\! H x\Bigr)
\ln\Bigl(1 \!+\! a H x\Bigr) \Biggr\} \!+\! 2 {\rm Li}_2\Bigl(
\frac{a \!+\! 1 \!+\! a H x}{2 a} \Bigr) \nonumber \\
& & \hspace{-.5cm} - 2 {\rm Li}_2\Bigl(\frac{a \!+\! 1 \!-\! a H
x}{2 a} \Bigr) \!+\! \Bigl(1 \!-\! \frac1{a} \!+\! H x\Bigr)
\ln\Bigl(1 \!-\! \frac1{a} \!+\! H x\Bigr) \ln\Bigl(\frac{a \!+\! 1
\!-\! a H x}{2}\Bigr) \nonumber \\
& & \hspace{-.5cm} + \Bigl(1 \!-\! \frac1{a} \!-\! H x\Bigr)
\Biggl\{ \ln\Bigl( \frac{a \!+\! 1 \!-\! a H x}{2 \!+\! 2 a H
x}\Bigr) \!-\! \ln\Bigl(1 \!-\! \frac1{a} \!-\! H x\Bigr) \ln\Bigl(
\frac{a \!+\! 1 \!+\! a H x}{2 \!+\! 2 a H x}\Bigr) \Biggr\} \nonumber \\
& & \hspace{3.8cm} + \frac2{a} \ln\Bigl( \frac{a \!+\! 1 \!-\! a H
x}{2}\Bigr) \!+\! 2 \ln(2) \ln\Bigl( \frac{a \!+\! 1 \!-\! a H x}{a
\!+\! 1 \!+\! a H x}\Bigr) \; , \qquad \\
\lefteqn{-2 x \mathcal{J}^{GK}_9 = (1 \!-\! \gamma) \Biggl\{ {\rm
Li}_2\Bigl( \frac{1 \!-\! a H x}2 \Bigr) \!-\! {\rm Li}_2\Bigl(
\frac12 \Bigr) \!+\!
{\rm Li}_2\Bigl(\frac{1 \!+\! a H x}{a \!+\! 1 \!+\! a H x}\Bigr) } \nonumber\\
& & \hspace{1.5cm} - {\rm Li}_2\Bigl( \frac{1 \!-\! a H x}{a \!+\! 1 \!-\!
a H x} \Bigr) \!-\! \ln(2) \ln\Bigl(1 \!+\! a H x\Bigr) \!+\! \frac12
\ln^2\Bigl( 1 \!+\! a H x\Bigr) \Biggr\} \; , \qquad \\
\lefteqn{-2 x \mathcal{J}^{GK}_{10} = 4\gamma \Biggl\{ - \ln\Bigl(1 \!+\!
a H x\Bigr) \!+\! \ln\Bigl( \frac{a \!+\! 1 \!+\! a H x}{a \!+\! 1 \!-\!
a H x}\Bigr) \Biggr\} \; . \qquad }
\end{eqnarray}
It remains just to give the expansions for large $a$ at fixed $x$,
and at fixed $r = a x$. In both cases the leading term goes like
$\ln(a)$. For fixed $x$ we find,
\begin{eqnarray}
-2 x \mathcal{J}^{GK}_1 & = & -4 \ln(2 H x) \ln(a) + O(1) \; , \\
-2 x \mathcal{J}^{GK}_2 & = & 0 \times \ln(a) + O(1) \; , \\
-2 x \mathcal{J}^{GK}_{34} & = & 4 \ln(Hx) \ln(a) + O(1) \; , \\
-2 x \mathcal{J}^{GK}_{56} & = & \Bigl[2Hx \!-\! 2 (1 \!-\! Hx)
\ln(1 \!-\! Hx)\Bigr] \ln(a) + O(1) \; , \\
-2 x \mathcal{J}^{GK}_{78} & = & \Bigl[2Hx \!+\! 2 (1 \!+\! Hx)
\ln(1 \!+\! Hx)\Bigr] \ln(a) + O(1) \; , \\
-2 x \mathcal{J}^{GK}_9 & = & 0 \times \ln(a) + O(1) \; , \\
-2 x \mathcal{J}^{GK}_{10} & = & -4 \gamma \ln(a) + O(1) \; .
\end{eqnarray}
At fixed $r$ the results are,
\begin{eqnarray}
-2 x \mathcal{J}^{GK}_1 & = & 0 \times \ln(a) + O(1) \; , \\
-2 x \mathcal{J}^{GK}_2 & = & -\Biggl[ \frac{\pi^2}{6} \!+\!
2 {\rm Li}_2\Bigl(\frac{Hr \!-\! 1}{Hr \!+\! 1}\Bigr) \Biggr] \ln(a)
+ O(1) \; , \\
-2 x \mathcal{J}^{GK}_{34} & = & -4 \ln(1 \!+\! Hr) \ln(a) + O(1) \; , \\
-2 x \mathcal{J}^{GK}_{56} & = & 0 \times \ln(a) + O(1) \; , \\
-2 x \mathcal{J}^{GK}_{78} & = & 0 \times \ln(a) + O(1) \; , \\
-2 x \mathcal{J}^{GK}_9 & = & 0 \times \ln(a) + O(1) \; , \\
-2 x \mathcal{J}^{GK}_{10} & = & 0 \times \ln(a) + O(1) \; .
\end{eqnarray}

\end{document}